\documentclass[useAMS,usenatbib]{mn2e}

\usepackage{psfig, epsf, epsfig}

\title[Globular cluster formation]{Globular cluster formation
with multiple stellar populations: self-enrichment in
fractal massive molecular  clouds}
\author[K. Bekki]
{Kenji Bekki${}^1$\thanks{E-mail:
kenji.bekki@uwa.edu.au} \\
${}^1$ICRAR M468
The University of Western Australia
35 Stirling Hwy, Crawley
Western Australia 6009, Australia}

\begin{document}

\date{Accepted, Received 2005 February 20; in original form }

\pagerange{\pageref{firstpage}--\pageref{lastpage}} \pubyear{2005}

\maketitle

\label{firstpage}

\begin{abstract}

Internal chemical abundance spreads are one of fundamental properties
of globular clusters (GCs) in the Galaxy. In order to understand the
origin of such abundance spreads, we numerically investigate
GC formation from massive molecular clouds (MCs) with fractal structures
using our new hydrodynamical simulations with star formation
and feedback effects of core-collapse supernovae (SNe) and asymptotic
giant branch (AGB) stars. We particularly investigate
star formation from gas chemically contaminated by SNe and AGB stars
(`self-enrichment') in forming GCs within MCs with different initial
conditions and environments. The principal results are as follows.
GCs with multiple generation of stars can be formed from merging
of hierarchical star cluster complexes that are developed from
high-density regions of fractal MCs.
Feedback effects of SNe and AGB stars can control the formation
efficiencies of stars formed from original gas of MCs
and from gas ejected from AGB stars.
The simulated GCs have strong  radial gradients of helium abundances
within the central 3 pc.
The original MC masses need to be as large as $10^7 {\rm M}_{\odot}$
for a canonical initial stellar mass
function (IMF)  
so that the final masses of stars formed 
from AGB ejecta can be  $\sim 10^5 {\rm M}_{\odot}$.
Since star formation from AGB ejecta is rather prolonged
($\sim 10^8$ yr), their formation
can be strongly suppressed by SNe of the stars themselves.
This result implies  that the so-called
mass budget problem is much more severe than ever thought in the 
self-enrichment scenario of GC formation
and thus  that IMF
for the second generation of stars should be
`top-light'.

\end{abstract}


\section{Introduction}

It has been established that internal chemical abundance spreads
are one of fundamental properties of old globular clusters
in the Galaxy (e.g., Gratton et al. 2012 for a recent review),
the Large and Small Magellanic Clouds (e.g., Mucciarelli et al. 2009;
Niederhofer et al. 2016),
and the Galactic dwarf satellites (e.g., Carretta et al. 2010;
Larsen et al. 2014 ).
Most old GCs in the Galaxy show internal chemical abundance spreads
in light elements (e.g., Carretta et al. 2009; C09)
whereas only 8 of them have been so far
observed to have internal [Fe/H] spreads (e.g., Marino et al. 2015).
NGC 2808 and $\omega$ Cen are GCs with He abundance spreads
(e.g., Piotto et al. 2005), the origin of which remains unclear.
The Galactic GC M22 is observed to 
have  at least two groups of stars with (i) the [Fe/H] difference
of $\sim 0.15$ dex among the two groups
and (ii) 
higher abundance of
$s$-process element in
the Fe-rich  group (e.g., Marino et al. 2009).
The origin of
the observed ubiquitous anti-correlations between light elements
and different levels of internal abundance spreads in GCs are
is one of unresolved problems in GC formation and evolution.

If these abundance spreads in various elements are due largely to
secondary star formation from gas contaminated by earlier generation
of stars within forming GCs (`self-enrichment'),
then we need to understand how such self-enrichment processes are
possible in such compact stellar systems of GCs.
Self-enrichment of pristine gas by AGB ejecta in forming GCs
is demonstrated to be
essentially important for the origin of the observed Na-O anticorrelations
among  GC stars (e.g., D'Ercole et al. 2010; D10).
Self-enrichment processes by SNe could explain the observed large metallicity
spread and metallicity distribution function in $\omega$ Cen
(e.g., Ikuta \& Arimoto 2000).
Lee et al. (2009) found possible evidence of Ca abundance spreads in 7 Galactic GCs 
and thus
suggested that self-enrichment processes by SNe are quite important
for the origin of the observed spreads in [Ca/Fe].

In spite of such importance of self-enrichment processes
in GC formation, only several numerical simulations of GC formation
have investigate the processes so far.
Bekki \& Chiba (2007) investigated how stellar wind of massive 
stars can influence the star formation processes and chemical evolution
of forming GCs 
within turbulent, high-density giant MCs.
They found that (i) second generation (`SG') of
 stars shows a C-N anticorrelation,
(ii) the observed high [N/Fe] of $\sim 0.8$ (e.g., NGC 6752)
can not be reproduced
in the simulated GCs for a canonical IMF,
and (iii) the fraction of SG stars formed from gas contaminated by
massive stars is quite small ($\sim 3$\%).
Using two-dimensional hydrodynamical simulations of star
clusters (SCs)
with stellar winds and SNe,
W\"unsch et al. (2008) performed 2D hydrodynamical simulations
of young SCs with supernova winds
and found  that a significant fraction of SN ejecta
can be still trapped in their inner regions if SCs are quite massive.
Bekki (2010, 2011; B10 and B11, respectively) 
demonstrated that star formation from AGB ejecta
can proceed very efficiently in clusters of 
first generation of stars (`FG'), as long as
the clusters are massive enough ($ \ge 10^6 {\rm M}_{\odot}$).

Although these previous simulations contributed to the better understanding
of self-enrichment processes of GCs,  the adopted initial conditions
and models for GC formation are quite idealized and less realistic
in the following points.
First, these simulations do not consider the observed 
fractarity of MCs (e.g., Blitz \& Williams 1990; Bergin \& Tafalla 2007). 
The fractal structures of MCs play key roles in the formation
and evolution processes of SCs (e.g., Elmegreen 2008), and the observed ubiquitous
SC complexes (e.g., Efremov 1995; Bastian et al. 2005; Adamo et al. 2012) can be developed
from such fractal structures and thus important for 
GC formation (Bekki 2017; B17).  Accordingly, the observed fractarity needs
to be included self-consistently
in a more sophisticated simulation of GC formation.
Second, feedback effects of SNe and AGB  winds are not simultaneously
and self-consistently included in previous simulations, which means that
self-enrichment processes are not so realistic: either only chemical
enrichment (and feedback effect) by SNe
or only that by AGB stars was included in previous simulations.
Accordingly, the previous models of GC formation did not predict
possible abundance spreads in heavy elements (due to chemical enrichment by  SNe)
and in light elements (AGB stars).

Third,  mass-dependent chemical yields of SNe and AGB 
are not properly included in previous chemodynamical simulations of GC formation.
Given that chemical yields are different
between SNe and AGB stars with different masses
(e.g., Karakas 2010; K10) and star formation can proceed within
a timescale of $10^6$ yr,
chemical abundance patterns of GC stars can depend strongly
on which SNe or massive AGB stars can contribute to chemical enrichment processes
within forming GCs. 
Accordingly time evolution of chemical abundances in ejecta of SNe and
AGB stars needs to be included self-consistently.
Fourth, secondary star formation from
gaseous ejecta from SNe and AGB stars  within an existing single giant
SC with the mass ($M_{\rm sc}$) larger than $10^6 {\rm M}_{\odot}$
is not so realistic, given that a SC is formed not in isolated but
as a group of smaller clusters
(e.g., Efremov 1995; Bastian et al. 2005).
Therefore,
self-enrichment processes investigated in previous 3D hydrodynamical simulations
of GC formation (e.g., D'Ercole et al. 2008, D08; B10, B11) in a gravitational potential
that is not evolving so much could be less realistic (B17)
Thus, more realistic initial conditions of GC formation
are required to be adopted so that self-enrichment processes of GCs
can be better investigated in numerical simulations of GC formation.

The purpose of this paper is to investigate self-enrichment
processes of  GCs formed within massive MCs with fractal structures
using new hydrodynamical simulations with feedback effects of 
SNe and AGB stars with different masses within MCs.
We consider that  GCs can be formed from massive MCs
with masses larger than $3 \times 10^6 {\rm M}_{\odot}$ 
within gas-rich
dwarf disk galaxies at high redshifts and thereby
investigate the transformation from fractal MCs into compact stellar 
systems (GCs) in detail.
We particularly investigate the following points: (i) how first generations
of stars can be formed from cold gas of fractal MCs,
(ii) whether new stars can be formed from gas ejected from SNe and AGB
stars during merging of hierarchical star cluster complexes developed
from fractal MCs,
and (iii) how feedback effects of SNe and AGB
winds influence the formation efficiencies
of FG and SG stars in GCs.
It should be noted here that  AGB winds can significantly influence
secondary star formation within forming GCs (Bekki 2016; B16).

The plan of the paper is as follows.
We describe the models for massive MCs with fractal structures,
feedback effects of SNe and AGB stars,
chemical enrichment by these stars,
star formation within MCs, and live gravitational  potentials of 
GC-host dwarf galaxies
in \S 2.
We present the key results of the simulation, in particular,
dynamics of GC formation from hierarchical star cluster complexes
developed from fractal MCs and self-enrichment processes in GC formation
in \S 3.
Based on these results,
we discuss (i) important roles of feedback effects 
of SNe and AGB stars in the physical properties
of GCs and (ii) possibly different IMFs between FG and SG star formation
in \S 4.
We summarize our  conclusions in \S 5.
The physical meanings of acronym and symbols (e.g., FG and SG)
often used in
this paper are summarized in Table 1 for convenience.

In the present paper, we consider that the origin of the observed abundance
spreads in the Galactic old GCs is due largely to multiple generations of 
stars in forming GCs. However, 
it is being hotly debated  whether the observed 
extended  main-sequence turn-offs (eMSTOs)  and splits of
main-sequence of the LMC clusters can result from age spreads
(i.e., multiple generation of stars)
or from internal stellar rotation (e.g.,
Bastian \&  De Mink 2009;  Milone et al. 2016; Li et al. 2016).
Accordingly, the above scenario of multiple  generation of stars in the Galactic GCs
could be just an assumption or hypothesis.
However, For \& Bekki (2017) have recently
discovered  young stellar objects (YSOs) with ages well less than
$10^6$ yr 
in  the older LMC SCs 
with ages of $0.1-1$ Gyr. This result 
is direct evidence for ongoing star formation in older LMC SCs
and therefore strongly suggests
that secondary star formation could have occurred in some of LMC SCs.
Therefore, the above scenario of multiple generation of stars in forming GCs
can be quite realistic, at least for some GCs, 
though the origin of the LMC SCs with multiple stellar populations 
can be different from that of GCs in galaxies other than the LMC.

\begin{table}
\centering
\begin{minipage}{85mm}
\caption{Description of (physical) meanings  for
acronym and symbols often used in the present study.}
\begin{tabular}{ll}
{ Acronym/Symbols } &
{ Physical meaning } \\
SC & Star cluster \\
MC  & Molecular cloud \\
SF  & Star formation \\
FG  & First generation of stars \\
SG  & Second generation of stars \\
$D_3$   & Fractal dimension (in 3D space) of a MC\\
$M_{\rm fg}$  & Total mass of FG stars \\
$M_{\rm sg}$  & Total mass of SG stars \\
$M_{\rm ns}$  & Total mass of new stars \\
$M_{\rm ej}$  & Gas mass ejected from SNe (AGB stars) \\
$\Sigma$  & Surface mass density (e.g., $\Sigma_{\rm g}$ for gas)  \\
$\epsilon_{\rm sf, fg}$  & Star formation efficiency of FG stars \\
$\epsilon_{\rm sf, sg}$  & Star formation efficiency of SG stars \\
$t_{\rm agb}$  & Lifetime of stars that become AGB stars. \\
$t_{\rm delay, sg}$  & Time delay between SF and SNe in SG. \\
$\rho_{\rm th}$  & Threshold gas density for SF. \\
\end{tabular}
\end{minipage}
\end{table}

\section{The model}

\subsection{An overview}

We consider that GCs can be formed from fractal MCs with
their initial masses ($M_{\rm mc}$)
much larger than the typical mass of the Galactic MCs
in gas-rich dwarf galaxies at high redshifts.
Harris \& Pudritz (1994) proposed that GC-hosting MCs should be
very massive (`super-massive MCs'), because star formation efficiencies
of MCs are typically rather low. 
The original mass of a MC ($M_{\rm mc}$) or a MC association
(a group of giant MCs) hosting
a GC with the initial mass of $M_{\rm gc, i}$ can be estimated
from the final GC mass ($M_{\rm gc, f}$)  by considering
(i) star formation efficiency within the MC ($\epsilon_{\rm sf}$),
(ii) gas ejection 
through SNe and AGB phases,
and (ii) mass loss due to dynamical evolution
(two-body relaxation and tidal stripping). 
The present-day (i.e., final)  mass of the GC
is  as follows:
\begin{equation}
M_{\rm gc, f}=(1-f_{\rm strip})(1-f_{\rm ej})M_{\rm gc,i},
\end{equation}
where $f_{\rm strip}$ is the mass fraction of stars lost 
from the GC due to dynamical evolution
and $f_{\rm ej}$ is the fraction of gas ejected from SNe and AGB stars.
Therefore, the initial mass of GC-hosing cloud is simply as follows:
\begin{equation}
M_{\rm mc}= \epsilon_{\rm sf}^{-1} M_{\rm gc, i}.
\end{equation}
For a typical mass of the Galactic GCs ($2 \times 10^5 {\rm M}_{\odot}$),
reasonable values of $f_{\rm ej}=0.4$ and $f_{\rm stri}=0.5$,
and rather high $\epsilon_{\rm sf}=0.2$,
$M_{\rm mc}$ can be therefore
$3.3 \times 10^6 {\rm M}_{\odot}$, which corresponds
to the most massive GMCs in the Galaxy (e.g., Solomon et al. 1979).
In this estimation, the initially rather large GC masses adopted
in previous self-enrichment scenarios 
for multiple generation of stars in GCs (e.g., D08;  B11)
are not considered. If such GC masses are considered, then
$M_{\rm mc}$ can be $[5-10]$ times larger than the above value.
Thus, we need to investigate GC formation in massive MCs with
$M_{\rm mc} \ge 3 \times 10^6 {\rm M}_{\odot}$ in order to discuss
the physical properties of GCs.

Giant molecular clouds (GMCs) are observed to have fractal structures
(e.g., Blitz et al. 2007),
and their origin and nature have been extensively discussed
both observationally and theoretically (e.g., Bergin \& Tafalla 2007;
Elmegreen 2008).
However it is not so clear how such fractal structures can influence
the formation processes of GCs, in particular, self-enrichment
processes that lead to the formation of SG stars $-$ the major component
of GCs. The key parameter of fractal MCs is the fractal dimension ($D_3$)
in three-dimensional (3D) space. Recent observations have shown that
$D_3$ in interstellar medium,  GMCs, and field stars  are different depending
on galaxy environments (e.g., Sun et al. 2016),
which implies that we need to choose reasonable ranges of $D_3$
depending on galaxy properties in the simulations of GC formation
within MCs.
By considering these observations, we investigate the influences of
initial fractal structures of MCs on GC formation.

In order to perform smooth particle hydrodynamics (SPH)
simulations 
of GC formation within massive MCs,
we use our own original simulation code that can be run
on GPU clusters (Bekki 2013, 2015). Although this code
enables us to investigate the formation of molecular hydrogen
(${\rm H_2}$) from neutral one on dust grains, dust formation,
destruction, and growth, effects of photo-electric heating
on cold gas,  star formation, SN feedback effects on star formation,
we do not include dust-related physics in the present simulation.
This is firstly  because we do not focus on dust physics 
in GC formation in the present study, and secondly because 
simulations with  such dust-related physics
are very time-consuming (Bekki 2015). Since the details of the code
are given in Bekki (2013, 2015), we briefly describe the code
in the present study.

\subsection{Massive molecular clouds}

We adopt a size-mass relation that is consistent with
(i) the observed relation between
mass densities and sizes of GMCs discovered
by Larson's (1981)
and (ii) the observed typical mass and size of GMCs in
the Galaxy (e.g., Solomon et al. 1979).
The following
$R_{\rm mc}-M_{\rm mc}$ relation is used for deriving
the size of a massive MC ($R_{\rm mc}$) from the mass ($M_{\rm mc}$)
for each MC;
\begin{equation}
R_{\rm mc}
 =40 \times  (\frac{M_{\rm mc}}{5 \times 10^5  {\rm M}_{\odot} })^{0.53}
{\rm pc}
\end{equation}
We investigate models with $M_{\rm mc}$
ranging from $3\times 10^5 {\rm M}_{\odot}$ 
to $10^7 {\rm M}_{\odot}$ in order to simulate
massive SCs (GCs) with the initial total masses larger than $10^5 {\rm M}_{\odot}$.
This wide range of investigation can allow us to derive
physical conditions for self-enrichment by AGB stars in a convincing manner.

A MC is assumed to have a power-law radial density profile
($\rho_{\rm mc}(r)$) as follows:
\begin{equation}
\rho_{\rm mc} (r)=\frac{\rho_{mc,0}}{ (r+c_{\rm mc})^{\beta} },
\end{equation}
where $r$,  $\rho_{\rm mc, 0}$,  and $c_{\rm mc}$,
$\beta$  are the distance from the MC's
center,  a constant that is determined by $M_{\rm mc}$ and $R_{\rm mc}$,
the core radius of the MC, and the power-law slope.
Although GMCs are observed to have $\beta = 1-2$ (e.g., Ashman \& Zepf 2001),
we consider that $\beta=1$ is more reasonable. This is because
the total mass of a GMC is roughly proportional to
$R^{3-\beta}$, for which $\beta=1$ is consistent with the above
mass-size relation ($M_{\rm mc} \propto R_{\rm mc}^2$).

A MC is assumed to have a fractal gaseous distribution  characterized by
a fractal dimension $D_3$. The details of a way to set up the initial
condition of a fractal structure for a given $\beta$ are given in 
Appendix A. In the present model for fractal MCs,  the power-law radial
density profile of a MC can be seen even in the smallest substructure
within the MC. Such a clumpy MC can show star formation in substructures
from the earlier evolution of the MC so that low-mass unbound and
bound SCs can be first formed from high-density regions of substructures.
Therefore, dynamics of GC formation in fractal MCs can be
significantly different from that in MCs without fractal structures. 
We consider that $D_3=2$ is more consistent with $\beta=1$,
because $M_{\rm mc}$ is scaled to $R_{\rm mc}^{D_3}$ $-$ a definition
of fractal dimension. We therefore investigate the models with
$D_3=2$ more extensively, though we also investigate
models with other $\beta$ for
comparison. 
The initial number of gas particles ($N_{\rm g}$)
used in a simulation depends on $D_3$ and 
$N_{\rm min}$, which is 
the minimum number of gas particles used in the Level 1 distribution of gas
particles 
(Appendix A).
It is initially 1048911 for the fiducial model (described later) and 
the total gas particle number
can increase with time owing to the ejection of gas from
AGB stars.

The initial virial ratio ($t_{\rm vir}$) can determine the total amount
of kinetic energy ($T_{\rm kin}$) of a MC and it is described as follows:
\begin{equation}
t_{\rm vir}=\frac{ 2T_{\rm kin} }{ | W_{\rm mc} | },
\end{equation}
where $W_{\rm mc}$ is the initial total potential energy of the MC.
The random motion of each gas particle is determined by 
the above equation for a given spatial distribution of a MC. 
We investigate mainly the models with $t_{\rm vir}=0.35$, because the formation
of compact stellar systems is ensured for that $t_{\rm vir}$ 
(e.g., Dale et al. 2014).
We present only the results of the models with $t_{\rm vir}=0.35$,
because other models with (i.e., $t_{\rm vir}=0.7$) show essentially similar behavior
in GC formation.
We also  consider
rigid rotation of a MC in some models, because previous observations suggested
that velocity gradients within MCs could be due to
such rotation (e.g., Phillips 1999; Rosolowsky et al. 2003). Since the magnitude of
rigid rotation is not so well constrained, we assume that
the amplitude of rigid rotation is a free parameter. Accordingly,
$T_{\rm kin}$ is the combination of the total random energy
$T_{\rm ran}$ and the total rotational one ($T_{\rm rot}$) as follows:
\begin{equation}
T_{\rm kin}=T_{\rm ran}+T_{\rm rot}.
\end{equation}
The way to give 3D velocities of gas particles based on
$T_{\rm ran}$ and $T_{\rm rot}$ is given in Appendix.

In order to discuss the importance of initial rotation
of MCs in GC formation,  we introduce the following parameter:
\begin{equation}
f_{\rm rot}=\frac{ T_{\rm rot} }{T_{\rm kin}}.
\end{equation}
We mainly discuss the results of the models with $t_{\rm rot}=0$,
and we show the results  for several rotating MC  models with $f_{\rm rot}=0.1$
in the present paper.
Single massive stellar systems can be
formed for such low $f_{\rm rot}$. The results of models with larger
$f_{\rm rot}$, for which binary clusters can be formed,
will be discussed in our forthcoming papers.
Initial gaseous temperature and metallicity are
set to be 10K and [Fe/H]=$-2$ in all MCs.
The radiative cooling processes
are properly included  by using the cooling curve by
Rosen \& Bregman (1995) for  $T < 10^4$K
and the MAPPING III code
for $T \ge 10^4$K
(Sutherland \& Dopita 1993).

\begin{table}
\centering
\begin{minipage}{85mm}
\caption{Description of key physical properties  for
the fiducial massive MC model.}
\begin{tabular}{ll}
{ Parameters } &
{ Values } \\
Initial MC mass   & $10^7 {\rm M}_{\odot}$  \\
Initial number of gas particles  & 1048911 \\
Mass resolution  &  $9.5 \times 10 {\rm M}_{\odot}$ \\
Size resolution  &  0.39 pc \\
Number of AGB particles per a gas particle  &  5\\
Number of SNe types (in mass)   &  4 \\
SN and AGB feedback from FG stars   & Yes \\
SN and AGB feedback from SG stars   & No \\
AGB yield  & K10 \\
SN yield  & T95 \\
Threshold gas density for star formation  & $10^4$ cm$^{-3}$ \\
Tidal field  from a dwarf host   & No \\
\end{tabular}
\end{minipage}
\end{table}

\subsection{Star formation and SN feedback}

Gas particles can be converted into collisionless new stellar particles
(`new stars') if 
the following two physical conditions can be met.
First  is that the local density ($\rho_{\rm g}$) exceeds a threshold density
(${\rho}_{\rm th}$) for
star formation:
\begin{equation}
{\rho}_{\rm g} > {\rho}_{\rm th}.
\end{equation}
We consider that star formation can proceed in the dense cores
of MCs, and accordingly, 
$\rho_{\rm th}$ is set to be $[10^4-10^5]$  H atoms cm$^{-3}$,
which is consistent with the observed values (e.g., Bergin \& Tafalla 2007).
Second  is that the local velocity field around a gas particle
is consistent with that for gravitationally collapsing,
which is formulated as follows
\begin{equation}
div {\bf v}<0 .
\end{equation}
One SPH gas particle is converted into just one new star in the present
study (i.e., not multiple times) so that the total particle number
can not dramatically increase during a simulation.

Each new star particle is born with a fixed IMF and an initial mass
$m_{\rm ns}$: it should be noted here that this $m_{\rm ms}$ is not
a mass of each individual star, which is denoted as $m_{\rm s}$.
The stellar mass decreases with  time owing to mass loss
by SNe Ia, SNe II, and, AGB stars and the final stellar mass
after  $\sim 3 \times 10^8$ yr evolution
(duration of a simulation)
can be significantly
different from $m_{\rm ns}$.
The mass loss from intermediate-mass and high-mass stars
($m_{\rm s} > 5 {\rm M}_{\odot}$) plays a significant role  in
self-enrichment processes of GC formation in the present study.
The adopted IMF in number is defined
as $\psi (m_{\rm s}) = C_i{m_{\rm s}}^{-\alpha}$,
where $m_{\rm s}$ is the initial mass of
each individual star and the slope $\alpha =2.35$
corresponds to the Salpeter IMF.
The normalization factor $C_i$ is a function of a stellar particle mass,
$m_{\rm l}$ (lower-mass cutoff), and $m_{\rm u}$ (upper-mass cutoff):
\begin{equation}
C_i=\frac{m_{ns}
\times (2-\alpha)}{{m_{\rm u}}^{2-\alpha}-{m_{\rm l}}^{2-\alpha}}.
\end{equation}
where $m_{\rm l}$ and $m_{\rm u}$ are  set to be   $0.1 {\rm M}_{\odot}$
and  $120 {\rm M}_{\odot}$, respectively.
Although we investigate only the models 
with  $\alpha=2.35$ in the present study,
the importance of top-heavy IMF in GC formation
will be discussed in our forthcoming papers
based on the results of the models with lower $\alpha$ (e.g., $\alpha=1.85$).

SNe of new stars can give thermal and kinematic perturbation
to their surrounding gas within GC-forming MCs. 
Each SN is assumed to eject the feedback energy ($E_{\rm sn}$)
of $10^{51}$ erg that is converted into thermal and kinetic
energy of gas surrounding SN.
Thornton et al. (1998)  investigated how much fraction of $E_{\rm sn}$
of a SN can be used for the increase of random motion of the surrounding gas
(`kinematic feedback'). In the present simulation,
multiple SN explosion can occur within a single MC at different epochs
within $\sim 30$ Myr after star formation  so that 
the energy-ratio of kinematic feedback to total SN energy ($f_{\rm kin}$) 
can be quite different from those predicted in previous simulations for single SN.
We consider that 
$f_{\rm kin}$ is rather high owing to interaction of expanding shells formed from
different SNe.

The way to distribute  $f_{\rm kin}E_{\rm sn}$
(i.e., kinetic feedback energy)  of SNe among neighbor gas particles
is described as follows. Each SN can eject gas with an initial ejection
speed of $v_{\rm ej}$, which is estimated from the following equation: 
\begin{equation}
f_{\rm kin}E_{\rm sn}=0.5 (m_{\rm s}-m_{\rm BH})v_{\rm ej}^2,
\end{equation}
where $m_{\rm BH}$ is the mass of a black hole that is left after SN explosion
for massive stars. The total mass of ejecta from a SN ($m_{\rm ej}$)
depends on $m_{\rm s}$ owing to different $m_{\rm BH}$.
For $m_{\rm s}=8 {\rm M}_{\odot}$ and $f_{\rm kin}=1$,
$v_{\rm ej}=3800$ km s$^{-1}$ ($m_{\rm ej}=6.5 {\rm M}_{\odot}$).
The kinetic energy of a SN is distributed equally among gas particles
surrounding the SN. If there are $N_{\rm nei}$ gas particles
around a SN, then $j$th gas particle can receive momentum of
$m_{\rm ej}v_{\rm ej}/N_{\rm nei}$ so that its velocity 
can be changed as follows: 
\begin{equation}
(m_{j}+m_{\rm ej})v_{j, k}^{'} = m_{j}v_{\rm j, k}+m_{\rm ej}v_{\rm ej},
\end{equation}
where $m_j$ is the mass of the gas particle before interaction with
the SN, $v_{j,k}$ and $v_{j,k}^{'}$ are the 3D velocity ($k=1,2,3$ 
correspond to $x$, $y$, and $z$ components of the velocity)
before and after gas-SN interaction, respectively.
Although different SNe with different initial $m_{\rm s}$ explode
at different times, 
we consider that one star formation event is followed by
the following four SN events for different stellar mass ranges:
$m_{\rm s} = [8-15] {\rm M}_{\odot}$,
$m_{\rm s} = [15-30]{\rm M}_{\odot}$,
$m_{\rm s} = [30-60] {\rm M}_{\odot}$,
and $m_{\rm s} = [60-120] {\rm M}_{\odot}$.
We adopt this model,
because it is very time-consuming
for the present study to change the 3D velocities
(and chemical abundances) of gas particles around
all SNe with different masses.

We consider that the time delay
between conversion of gas into a
new star and a SN explosion 
is parameterized by $t_{\rm delay}$, which is describe as follows:
\begin{equation}
t_{\rm delay}=t_{\rm to}+t_{\rm sf},
\end{equation}
where $t_{\rm to}$ is the main-sequence turn-off timescale
and $t_{\rm sf}$ is the timescale of a pre-main sequence phase.
Since the present simulation can not resolve the formation
of individual stars from collapsing 
MC cores,  
we can not directly derive the timescale of a pre-main sequence
phase.
Given that the observed age of young stellar objects (YSOs)
is $\sim  10^6$ yr for massive stars (e.g., Whitney et al. 2008),
$t_{\rm sf}$ could be at least $10^6$ Myr.
This $t_{\rm sf}$ 
is negligibly short in comparison with the main-sequence
timescales of stars. 
We therefore investigated only the models with  $t_{\rm sf}=0$ 
in the present study. It should be noted
here that  $t_{\rm sf}$ can be quite long for low-mass
stars. However, inclusion of such long
$t_{\rm sf}$ for low-mass stars in the present simulations would not
change the present results significantly, because energetic feedback effects
from low-mass stars on interstellar medium (ISM) are not possible.

In order to calculate $t_{\rm to}$
from the main-sequence
turn-off mass ($m_{\rm to}$),
we use the following formula
(Greggio \& Renzini 2011):
\begin{equation}
\log m_{\rm to}(t_{\rm s})
= 0.0434 (\log t_{\rm s})^2 - 1.146 \log t_{\rm s} + 7.119,
\end{equation}
where $m_{\rm to}$  is in solar units and time $t_{\rm s}$ in years.
Using the above equation, we can derive $t_{\rm to}$
for a given $m_{\rm to}$ ($=m_{\rm s}$).
This is not a good approximation only for massive stars 
($m_{\rm s} > 10 {\rm M}_{\odot}$ that explode as SNe;
Greggio \& Renzini 2011). For such massive stars,
we adopt $m_{\rm s}=3.0 \times 10^7$ yr, $1.0 \times 10^7$ yr,
$4.9 \times 10^6$ yr, and $3.4 \times 10^6$ yr
for SNe with 
$m_{\rm s} = [8-15] {\rm M}_{\odot}$,
$m_{\rm s} = [15-30]{\rm M}_{\odot}$,
$m_{\rm s} = [30-60] {\rm M}_{\odot}$,
and $m_{\rm s} = [60-120] {\rm M}_{\odot}$,
respectively.
The average SN explosion time ($t_{\rm sn}$) of these four discrete
SN groups for the Salpeter IMF
is $1.42 \times 10^7$ yr.

Although time delay between star formation and SN explosion 
($t_{\rm delay, fg}$) is considered  to be dependent on $m_{\rm s}$
for all FG stars,  
we adopt a different model for time delay
between star formation of SG stars and SN explosion ($t_{\rm delay, sg}$).
This is because SG stars can be formed from AGB ejecta very efficiently
in B10 and B11 in which SNe were not included at all (i.e., no SN feedback
effects).
We investigate how the SN explosion from  SG stars can influence 
the formation processes of GCs with models with different $t_{\rm delay, sg}$.
If the upper-mass cutoff of the IMF ($m_{\rm u}$) is
lower, then $t_{\rm delay, sg}$ can be longer. By changing 
$t_{\rm delay, sg}$, we can discuss how the IMF of SG stars
can control the physical properties of SG stars, which are the
main components of GCs. We investigate the models without
SNe from SG stars and those with $t_{\rm delay, dg}=3$ Myr,
10 Myr, and 30 Myr.

\begin{table*}
\centering
\begin{minipage}{160mm}
\caption{The basic model parameters for
fractal molecular clouds (MCs).}
\begin{tabular}{llllllll}
{ Model ID } &
{ $M_{\rm mc}$ \footnote{ The initial total mass of 
a fractal molecular cloud (MC) in units of $10^6 {\rm M}_{\odot}$.
 }} & 
{ $R_{\rm mc}$ \footnote{ The initial size for a MC
in units of pc.
 }} & 
{ $f_{\rm rot}$  \footnote{ The initial ratio of total rotational
energy to total kinetic energy in a MC.
}} & 
{ $D_3$  \footnote{  The 3D fractal dimension of a MC.
}} &
{ $t_{\rm delay,sn}$  \footnote{ The time delay between
star formation  and the explosion of SNe in the formation of SG stars
(Myr).
The symbol `-' means that  no SNe can originate from SG stars
owing to a top-light IMF in the model.
}} &
{ $r_{\rm p}$  \footnote{ The initial position
of a MC with respect to the center of
its host galaxy in unit of kpc.
The symbol `-' means that the model does not include
the live gravitational potential of the MC-host galaxy.
}} &
{ comments } \\
M1 & 10 & 200 & 0 & 2 & - & - & fiducial \\
M2 & 10 & 200 & 0 & 2 & 3 & - &  \\
M3 & 10 & 200 & 0 & 2 & 10 & - &  \\
M4 & 10 & 200 & 0 & 2 & 30 & - &  \\
M5 & 10 & 200 & 0 & 2 & 30 & - &  no SN feedback for FG an SG\\
M6 & 10 & 200 & 0.1 & 2 & - & - & rotating MC \\
M7 & 10 & 200 & 0 & 2.4 & - & - & larger fractal dimension  \\
M8 & 10 & 200 & 0 & 2.4 & 3 & - &  \\
M9 & 10 & 200 & 0 & 3 & - & - &   large fractal dimension\\
M10 & 10 & 200 & 0 & 2 & - & - & $\rho_{\rm th}=10^5$ cm$^{-3}$ \\
M11 & 10 & 200 & 0 & 2 & 3 & - & $\rho_{\rm th}=10^5$ cm$^{-3}$ \\
M12 & 10 & 200 & 0 & 2 & - & - &  $\rho_{\rm mc}(r) \propto r^{-2}$\\
M13 & 10 & 200 & 0 & 2 & 3 & - &  $\rho_{\rm mc}(r) \propto r^{-2}$\\
M14 & 10 & 290 & 0 & 2 & - & - &  lower gas density \\
M15 & 10 & 139 & 0 & 2 & - & - &  higher gas density \\
M16 & 10 & 139 & 0 & 2 & 3 & - &   \\
M17 & 10 & 200 & 0 & 2 & - & 0.3 &  tidal field of MC-host galaxy \\
M18 & 10 & 200 & 0 & 2 & - &  1.0&  \\
M19 & 3 & 100 & 0 & 2 & - & - &  \\
M20 & 3 & 100 & 0 & 2 & 3 & - &  \\
M21 &  3 & 100 & 0.1 & 2 & - & - &  \\
M22 &  3 & 100 & 0.1 & 2 & 3 & - &  \\
M23 & 3 & 100 & 0 & 2.4 & - & - &  \\
M24 & 3 & 100 & 0 & 2.4 & 3 & - &  \\
M25 & 3 & 69 & 0 & 2 & - & - &  \\
M26 & 3 & 69 & 0 & 2 & 3 & - &  \\
M27 & 3 & 100 & 0 & 2 & - & - &  $\rho_{\rm th}=10^5$ cm$^{-3}$ \\
M28 & 3 & 100 & 0 & 2 & 3 & - &  $\rho_{\rm th}=10^5$ cm$^{-3}$ \\
M29 & 3 & 100 & 0 & 2 & - & 0.3 &  \\
M30 & 3 & 100 & 0 & 2 & - & 1.0 &  \\
M31 & 1 & 68 & 0 & 2 & - & - &  \\
M32 & 1 & 68 & 0 & 2 & 3 & - &  \\
M33 & 1 & 68 & 0 & 2 & - & - &  $\rho_{\rm th}=10^5$ cm$^{-3}$ \\
M34 & 1 & 68 & 0 & 2.4 & - & - &  \\
M35 & 1 & 68 & 0 & 2.4 & 3 & - &  \\
M36 & 0.3 & 32 & 0 & 2 & - & - &  \\
M37 & 0.3 & 32 & 0 & 2 &  3 & - &  \\
M38 & 0.3 & 32 & 0 & 2 & - & - &  $\rho_{\rm th}=10^5$ cm$^{-3}$ \\
\end{tabular}
\end{minipage}
\end{table*}

\subsection{Gas ejection and feedback effects from AGB stars}

Since AGB stars can eject gas significantly later than SNe,
SN explosion can expel almost all of original cold gas around 
intermediate-mass stars.
Accordingly, there can be almost no gas around the stars when
they start to eject gas during AGB phases. This is a serious problem
in implementing chemical enrichment of gas by AGB ejecta if
we adopt a standard model of chemical enrichment
in which chemical abundances of gas particles can change only
when the particles are within a certain radius from an AGB star: if no
gas particles around an AGB star, the AGB ejecta can not be given
to any particles (no chemical enrichment).
This unrealistic situation needs to be avoided in the present simulation
in which chemical enrichment processes are investigated.
We therefore adopt a novel model (B16) in which each AGB star
eject gas particles 
with chemical abundances predicted from recent AGB models
(e.g., K10). 
Ejection of new particles from AGB stars (`AGB particle') means that
the total number of gas particles can significantly increase as a simulation
goes.  

Chemical abundances of light elements are quite different between
AGB stars with different masses (e.g., K10; Ventura et al. 2013),
which means that SG stars formed from AGB ejecta at different times 
can have different abundances of light elements.
In order to model chemical enrichment by AGB stars with different masses
more properly, we consider ejection of AGB particles at 
different five epochs ($t_{\rm agb}$, which corresponds
to $t_{\rm delay}$ for intermediate-mass stars).
These five are 200, 120, 80, 60, and 40 Myr
and correspond to the lifetimes of
the  masses of
stars,
(i) $3 \le m_{\rm s} < 4$ (${\rm M}_{\odot}$),
(ii) $4 \le m_{\rm s} < 5$ (${\rm M}_{\odot}$),
(iii) $5 \le m_{\rm s} < 6$ (${\rm M}_{\odot}$),
(iv) $6 \le m_{\rm s} < 7$ (${\rm M}_{\odot}$),
and (v) $7 \le m_{\rm s} <8$ (${\rm M}_{\odot}$),
respectively.
Here we consider only five different AGB particles,
because we can not use excessively large number of gas particles
owing to the limited amount of simulation time allocated to the project
of the present study.
The minimum mass of AGB stars ($3 {\rm M}_{\odot}$) is chosen,
firstly because
age differences between FG and SG stars can not be too large,
and secondary because the fraction of gaseous ejecta from 
AGB stars with $m_{\rm s}<3 {\rm M}_{\odot}$ is small.

An AGB particle is ejected from a new stellar particle
with a wind velocity of $v_{\rm wind}$
at the end of the  main-sequence phase of the stellar particle.
Although this $v_{\rm wind}$ is an order of 10 km s$^{-1}$,
such stellar wind can dramatically influence the star formation
histories within existing SCs (e.g., D08, B10, B11, and B16): such stellar
wind can be equivalent to kinematic feedback effects of SNe.
We adopt $v_{\rm wind}= 10$  km s$^{-1}$, which is
consistent with
recent observations of AGB stars in the LMC (e.g., Marshall et al. 2004).
The initial temperature of AGB wind ($T_{\rm wind}$)  is set to be 1000 K,
which is consistent with standard theoretical  models of  AGB winds. 
It is likely that SG formation from gas is possible only
if stellar wind can be efficiently cooled down from $T_{\rm wind}$
to a few tens K.

If the wind velocities  of AGB stars in a proto-GC
exceed the escape velocity ($v_{\rm esc}$) of the proto-GC, then 
the AGB ejecta is likely to escape from the proto-GC.
This condition is simply described as follows:
\begin{equation}
v_{\rm wind} > v_{\rm esc}= f(M_{\rm mc}, R_{\rm mc}, D_3),
\end{equation}
where  $v_{\rm esc}$ is a function of 
$M_{\rm mc}$, $R_{\rm mc}$, and $D_3$,
all of which can determine the gravitational potential of the FG stellar systems. 
As shown in B11,
low-mass FG systems are unlikely to retain AGB ejecta, because the above
condition is not satisfied.
Also,
in order to understand the importance of AGB feedback effects on
SG star formation, we investigated models with $v_{\rm wind}=0$
and those with $T_{\rm wind}=10$ K for comparison.
We confirmed that both $M_{\rm sg}$ and $\epsilon_{\rm sf, sg}$ are
higher in those models without AGB feedback effects, which is consistent
with the results in B11 and B16. Since this result appears to be obvious
(initially expected), we do not discuss these in this paper.

\begin{figure*}
\psfig{file=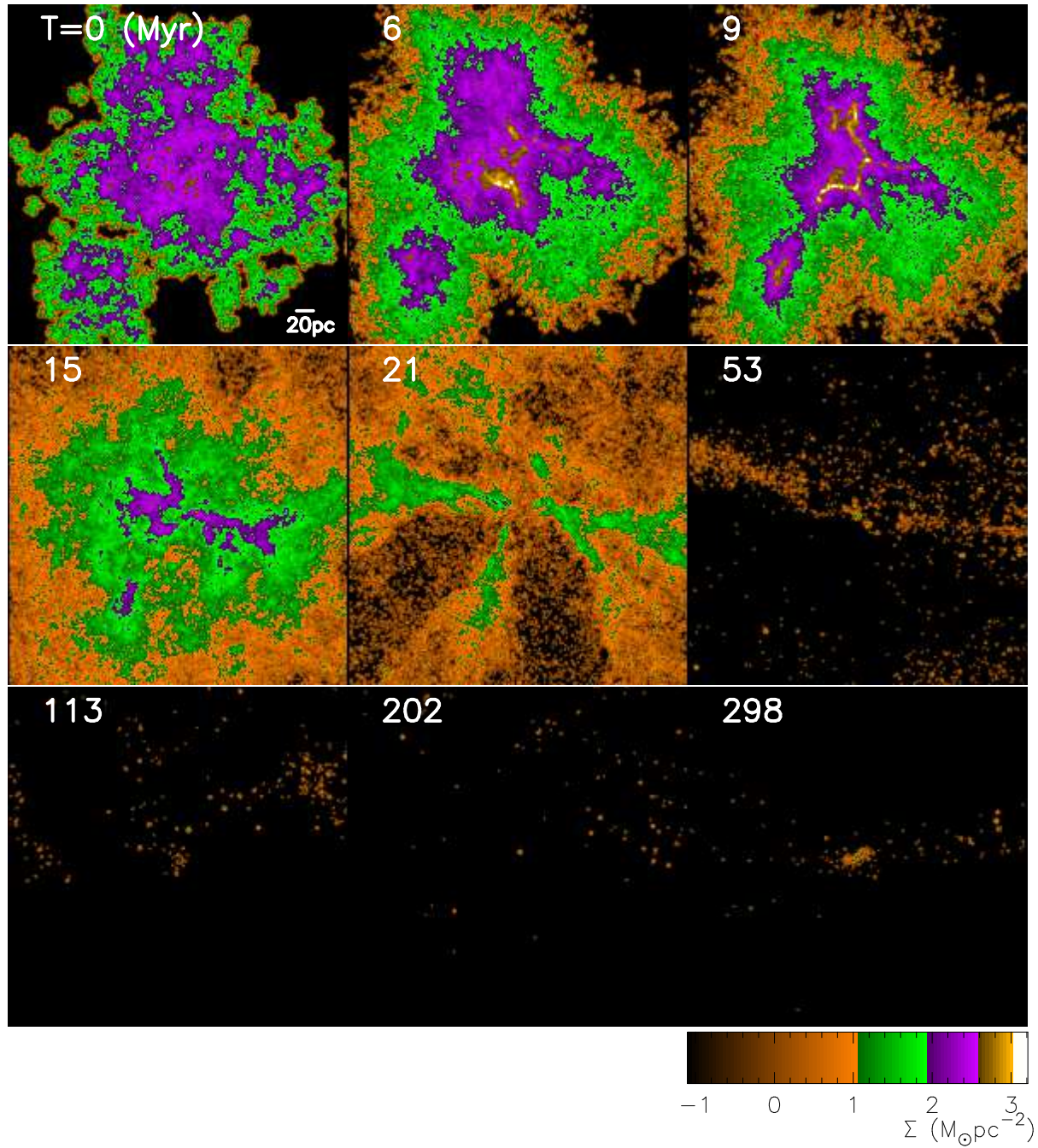,width=16.0cm}
\caption{
Time evolution of the surface mass density ($\Sigma$)  of original pristine gas
projected onto the $x$-$y$ plane for the fiducial model
with $M_{\rm mc}=10^7 {\rm M}_{\odot}$, $R_{\rm mc}=200$ pc,
and $D_3$=2. The time $T$ at the upper left conner in each frame
is given in units of Myr.  A thick bar in each panel indicates
a scale of 20pc.
For clarity, a color code in this figure is different from those used in
Figs. 2, 3, and 4.
}
\label{Figure. 1}
\end{figure*}

\begin{figure*}
\psfig{file=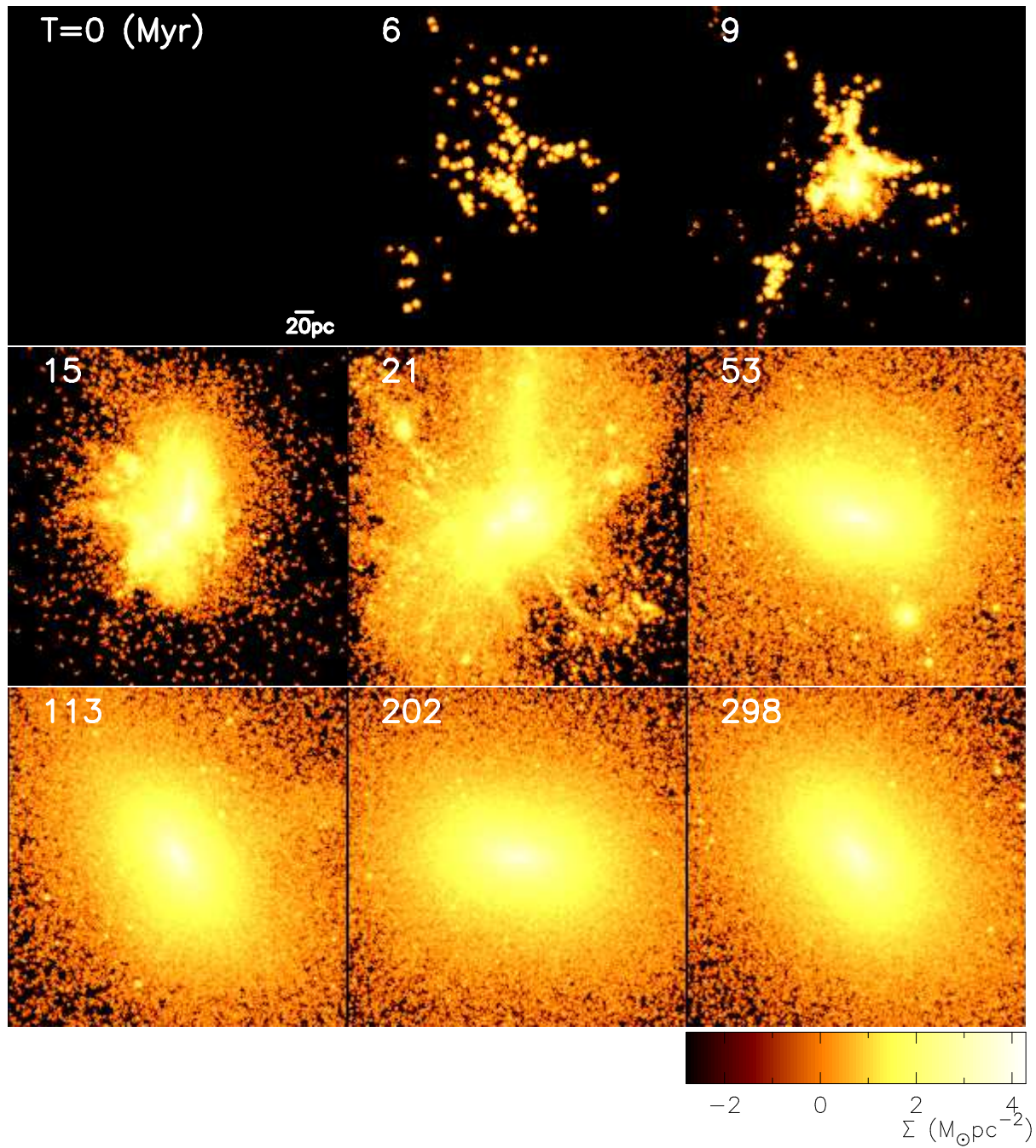,width=16.0cm}
\caption{
The same as Fig. 1 but for new stars formed from original gas (`FG' stars).
}
\label{Figure. 2}
\end{figure*}

\begin{figure*}
\psfig{file=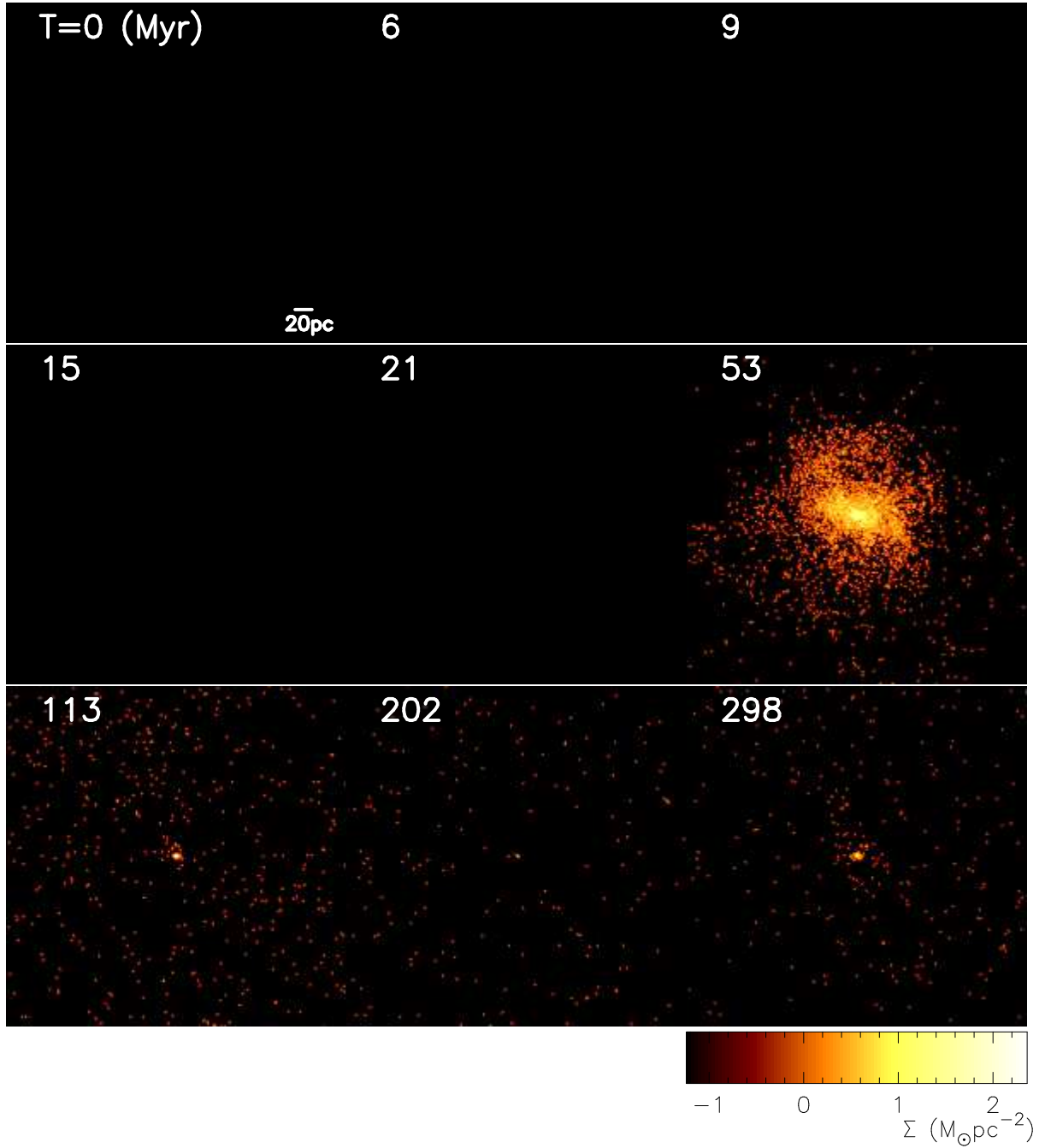,width=16.0cm}
\caption{
The same as Fig. 1 but for gas ejected  from AGB stars of FG.
}
\label{Figure. 3}
\end{figure*}

\begin{figure*}
\psfig{file=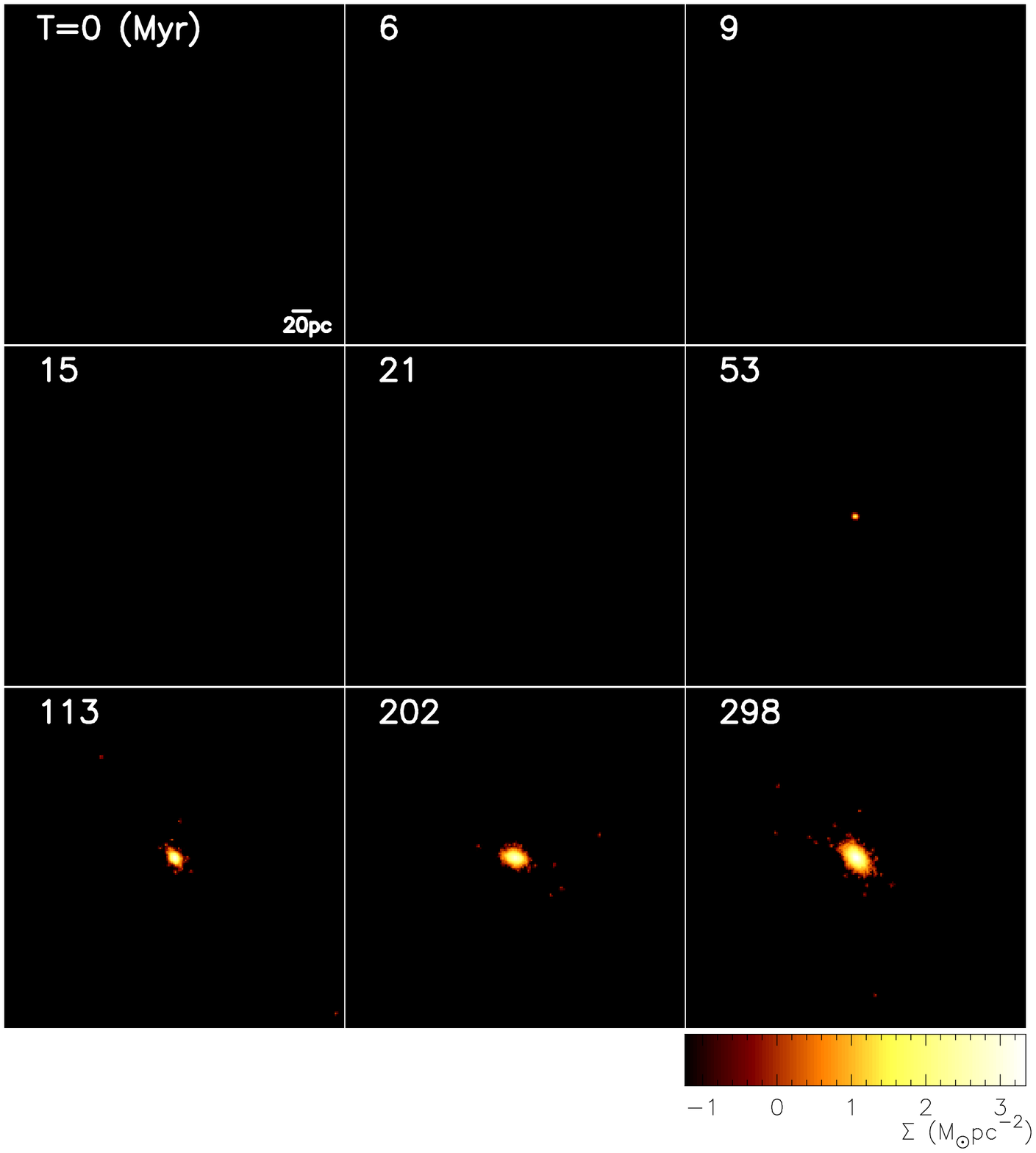,width=16.0cm}
\caption{
The same as Fig. 1 but for new stars formed from AGB ejecta of FG (`SG' stars).
}
\label{Figure. 4}
\end{figure*}

\subsection{Chemical enrichment}

Gaseous ejecta from  a SN can mix with its surrounding gas particles
so that the gas particles can increase their chemical abundances.
We consider that such increment can occur if the gas particles
are located 
within $r_{\rm sn}$ from the SN.
This $r_{\rm sn}$ is set to be the initial gravitational 
softening length (0.4 pc). The  chemical abundance of $k$th element
($k$=1, 2, 3,.... correspond to H, He, C, N, O. ... respectively)
for $j$th gas particle ($Z_{j, k}$) among $N_{\rm sn}$ surrounding
gas particles around a SN
can change according to the following equation:
\begin{equation}
(m_{j}+m_{\rm ej}) Z_{j,k}^{'} = m_{j}Z_{j,k}+ 
\frac{ \Delta m_{\rm ej} Z_{\rm sn, \it k} }{ N_{\rm sn} },
\end{equation}
where $Z_{j,k}^{'}$ are the chemical abundance of $k$th element
after chemical enrichment by the SN
and $Z_{\rm sn, \it k}$ is the
chemical abundance of $k$th element for the SN ejecta.
We use the chemical yield table of SNII from Tsujimoto et al. (1995, T95)
to calculate $Z_{\rm sn,k}$
in the present study.
Since we will describe the chemical
abundances of GC stars with different ages and locations
within GCs and their dependencies on model parameters in our next paper,
we briefly show some of the result in the present paper.

\subsection{Live galactic potential}

The tidal field of a dwarf galaxy hosting MCs can influence
the formation processes of GCs within MCs during the orbital evolution
of MCs around the MC-host dwarf.
We therefore investigate such tidal effects on GC formation
by  constructing  a model for live galactic potential
of a dwarf galaxy  as follows.
We assume that a MC-host dwarf galaxy 
consists of a dark matter halo
and a stellar disk.
Each of these components is
represented by collisionless N-body particles in the present study:
the galactic potential is `live' so that not only
tidal effect of a MC-host dwarf on a MC but also dynamical friction
of a MC against disk field stars of the dwarf can be self-consistently
included. Therefore, the present study is more sophisticated and
more realistic than our previous simulations of GC formation 
under a fixed galactic potential (e.g., Hurley \& Bekki 2007;
Bekki \& Chiba 2007).

The present model for a dwarf galaxy  is purely collisionless one,
which means that
gas dynamics,  star formation, chemical evolution,
and dust formation and evolution
are not included at all,
though the present simulation code enables us to investigate
these physical processes.
The dark matter halo 
with the total mass of $M_{\rm h}$ is represented by
the `NFW' one (Navarro et al. 1996)
with a central cusp predicted by the Cold Dark Matter (CDM)  model:
\begin{equation}
{\rho}(r)=\frac{\rho_{0}}{(r/r_{\rm s})(1+r/r_{\rm s})^2},
\end{equation}
where $r$,  $\rho_{0}$,  and $r_{\rm s}$ are the distance from the center
of the cluster, the central density, and the scale-length of the dark halo,
respectively.
The virial radius ($r_{\rm vir}$),  the scale radius ($r_{\rm s}$),
and the `$c$' parameter (=$r_{\rm vir}/r_{\rm s}$)
are chosen such that the values
are consistent with recent cosmological simulations 
for the adopted $M_{\rm h}$
(Neto et al. 2007).

The dwarf is assumed to be as a  bulge-less disk galaxy
with the total stellar mass of $M_{\rm s}$ and the size of $R_{\rm s}$.
The radial ($R$) and vertical ($Z$) density profiles of the stellar disk are
assumed to be proportional to $\exp (-R/R_{0}) $ with scale
length $R_{0} = 0.2R_{\rm s}$ and to ${\rm sech}^2 (Z/Z_{0})$ with scale
length $Z_{0} = 0.04R_{\rm s}$ , respectively.
In addition to the
rotational velocity caused by the gravitational field of disk
and dark halo components, the initial radial and azimuthal
velocity dispersions are assigned to the disc component according to
the epicyclic theory with Toomre's parameter $Q$ = 1.5.  The
vertical velocity dispersion at a given radius is set to be 0.5
times as large as the radial velocity dispersion at that point.

We investigate only one dwarf model in this study, because
the main purpose of this paper is to investigate not
the dynamical influences of MC-host dwarfs
with different masses and types on  star-forming MCs but the GC formation in
fractal MCs. The dwarf galaxy is assumed
to have 
$M_{\rm h}=10^{10} {\rm M}_{\odot}$,
$M_{\rm s}=6.0 \times  10^{7} {\rm M}_{\odot}$,
$R_{\rm s}=1.8$ kpc,
and no gas.
The mass and size resolutions for the simulation of the dwarf are
$6 \times 10^2 {\rm M}_{\odot}$ and 25 pc, respectively.
These resolutions are much lower than those
for a MC with $M_{\rm mc}=10^7 {\rm M}_{\odot}$ 
($9.5 \times 10 {\rm M}_{\odot}$ and 0.4 pc, respectively).
In order to avoid unrealistically strong gravitational influences  of clumpy
distributions of dark matter and disk particles of a dwarf
on the evolution of its  MC,
we adopt (i) multiple gravitational softening methods (Bekki \& Tsujimoto 2016)
and (ii) minimum time step width for numerically integrating
different equations for the dwarf's particles being
the same as that for the MC.

The initial position ({\bf x})  of a MC orbiting around its host dwarf is only a parameter
for the tidal effects of MC-host dwarfs in the present study. The 3D 
position ($x$, $y$, and $z$) of a MC within a dwarf  is given as follows:
\begin{equation}
{\bf x} = (r_{\rm p}, 0, 0),
\end{equation}
where $r_{\rm p}$ is the distance of the MC from the dwarf's center. The MC is 
assumed to have a circular motion  within the dwarf's
disk plane  initially. Therefore, its 3D
velocity ({\bf v}) is given as follows:
\begin{equation}
{\bf v} = (0, v_{\rm c}, 0),
\end{equation}
where $v_{\rm c}$ is the circular velocity at the position {\bf x}. 
Accordingly, $v_{\rm c}$ is determined by the adopted live galactic potential
of the dwarf.
We investigate the models with $r_{\rm p}=0.3$ and 1 kpc in the present study.

\subsection{Parameter study}

The key parameters for GC formation from fractal MCs  are
$M_{\rm mc}$, $R_{\rm mc}$, $f_{\rm rot}$, and  $D_3$,
for a given  IMF slope.  Although other parameters such as initial
radial density profiles of MCs
and galactic potentials are also important,
but we do not intend to discuss much about the roles of these
parameters in GC formation.
We mainly describe the results of the fiducial model in which
$M_{\rm mc}=10^7 {\rm M}_{\odot}$,
$R_{\rm mc}=200$ pc,
$f_{\rm rot}=0$, and
$D_{\rm 3}=2$,
because this model shows the typical behavior of GC formation with
multiple generations of stars within fractal MCs.
The basic parameters used for the fiducial model is summarized in Table 2.
We also discuss the results of other models with different values
of the key parameters. The parameter values of all 38 models discussed
in this paper are summarized in Table 3.

We focus mainly on the physical properties
of the simulated GCs in models without SNe from SG stars, because
it is expected that the total masses of SG stars ($M_{\rm sg}$) are
unlikely to be as large as $10^5 {\rm M}_{\odot}$ 
owing to SF suppression by SNe in the models with SNe from SG stars (B11).
However, we extensively investigate how SNe from SG stars can influence final
$M_{\rm sg}$ of GCs 
using models with SNe from SG stars.
We do not discuss the chemical abundances of simulated GCs 
so extensively in the present study,
because the present paper already contains a substantial amount of 
new results and it is long. We accordingly discuss the chemical properties
of the simulated GCs in our next paper using
AGB and SN yields not only from
K10 and T95
but also from other groups (e.g., Ventura et al. 2013).
The results of models with ISM of dwarfs (i.e, possible gas that can dilute
AGB ejecta) will be discussed in our forthcoming papers too.

\begin{figure}
\psfig{file=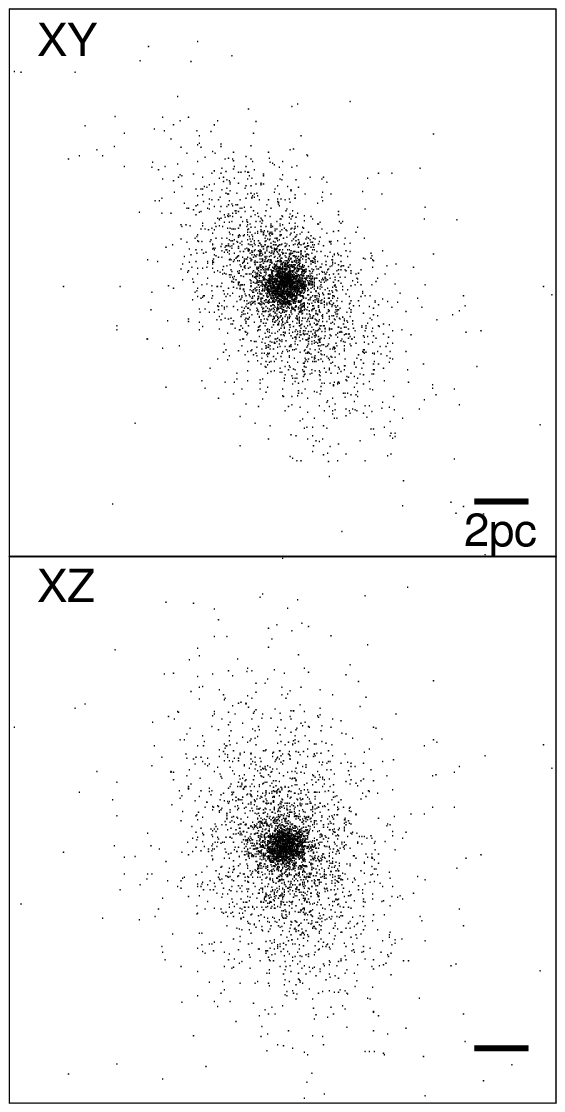,width=8.0cm}
\caption{
Mass distributions of SG stars projected onto the $x$-$y$ (upper)
and $x$-$z$ planes (lower) at $T=113$ Myr in the fiducial model M1.
A thick bar indicates a scale of 2 pc.
}
\label{Figure. 5}
\end{figure}

\section{Results}

\subsection{Dynamics of two-stage GC formation}

Figs. 1-4 show how a massive GC consisting of FG and SG stars
can be formed from a fractal MC in the fiducial
model M1 with $M_{\rm mc}=10^7 {\rm M}_{\odot}$ and $R_{\rm mc}=200$ pc
and without SN feedback effect for stars formed from AGB ejecta.
Numerous small gas clumps
can be  developed from local gravitational instability  within the MC,
and their  local gas densities can become higher than $10^4$ cm$^{-3}$
($T=6$ Myr).
As a result of this,
new stars can form in these gravitationally bound low-mass clumps 
($T=9$ Myr) to become new low-mass SCs.
These new stars formed from pristine low-metallicity cold gas
correspond to FG stars.
The new SCs can merge with one another within the MC  to form
a single FG stellar system over the timescale of $10^8$ yr.
Star formation can proceed also in massive long filaments developed during
the dynamical evolution of the fractal MC.

Multiple SN  explosion can occur well before most of the cold gas
is consumed by star formation, 
because SN explosion of massive stars with shorter lifetimes
($< 10$ Myr corresponding to $m_{\rm s} > 30 {\rm M}_{\odot}$)
are included in this fiducial model.
Consequently, a significant fraction of cold gas that was not converted into
new stars before SN explosion can be brown away from the MC. 
Once cold gas is  expelled from the MC ($T=21$ Myr),
most of the gas  can never be returned back to the inner part
of the MC owing to the shallow gravitational potential
(and to the non-inclusion of galactic potential).
In the present fractal MC model,
SN explosion can  occur during merging of low-mass SCs,
which is in striking contrast with SN explosion in the uniform distribution
of cold gas in MCs. 
About 46\% of initial cold gas can gain a large amount
of momentum and be heated up so that the gas can be finally completely removed
from the MC through SN explosion in this model.
The final star formation efficiency of FG stars  
($\epsilon_{\rm fg} \sim 0.5 $) in this model  is higher than $0.2-0.5$ required
for the formation of bound SCs (e.g., Hills 1980). 

Chemical enrichment of pristine gas by SNe can proceed during multiple merging
of low-mass SCs and new stars can be formed from such chemically polluted
gas ($T=15, 21$ Myr). However, 
the fraction of the new stars is quite small,
because almost all of the chemically polluted  gas can be removed from the proto-GC
region ($R<100$ pc) during GC formation. 
Furthermore such new star formation can occur well outside the inner region
of the forming GC
so that they can not be finally within the central region of the SG stellar system
(i.e., the main component of the GC) later formed.
Thus it is not possible  that the simulated GC can have significant 
internal abundance spreads in heavy elements (i.e., $\delta$[Fe/H]$>0.05$ dex)
between their stars within the central 5 pc.

After the removal of gas chemically polluted by SN explosion,
massive AGB stars ($m_{\rm s} =[7-8] {\rm M}_{\odot}$) start to eject
Na-rich (He-rich) gas into the MC ($T=53$ Myr).
Because of the relatively slow wind velocity ($v_{\rm wind}=10$ km s$^{-1}$),
the AGB ejecta can be gravitationally trapped in the central region
of the MC where a massive stellar system composed of FG stars only
is developing ($T=53$ Myr). The AGB ejecta can be slowly accumulated in
the inner region of the FG stellar system and finally converted into
new stars when the gas density exceeds $\rho_{\rm th}$.
These new stars are SG in the MC
and have a very compact spatial distribution initially.
As shown in our previous works (B10, B11), this compact configuration
is due largely to energy dissipation of gas during its accretion process.
The SG stellar system is initially composed mostly of new stars formed from
gas ejected from massive AGB stars with $m_{\rm s} \ge 6 {\rm M}_{\odot}$.
The SG system grow slowly by accretion of gas from AGB stars with
lower $m_{\rm s}$ over a timescale of $10^8$ yr. New SG stars formed
from ejecta from AGB stars with lower masses can be distributed in the
outer part of the SG system, which implies that He-rich SG stars 
has a more compact distribution than He-normal ones in a GC.

\begin{figure}
\psfig{file=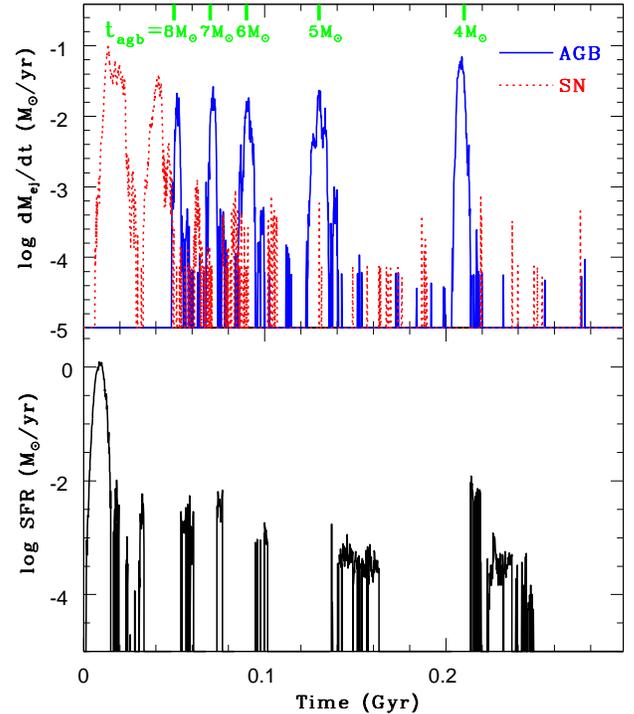,width=9.0cm}
\caption{
Time evolution of ejection rates of gas from SNe (red dotted) and from
AGB stars (blue solid) in a proto-GC (upper) and the star formation history
of the GC (lower) for the fiducial model. 
The epochs of gas ejection
from AGB star ($t_{\rm agb}$) with different masses are shown by thick green lines 
at the top of the upper frame for the initial star burst around $T=5$ Myr
in this model.
Chemical pollution
and feedback effects by SNe from SG stars are not included in this model.
Therefore, all of SNe in this figure are from FG stars.
Although most of the original gas can be expelled from the forming
GC by $T=0.04$ Gyr, 
a very minor fraction of the gas can settle down to the inner region
of the forming GC at later times, because the gas is not influenced by
SN feedback effects (Only gas that is not influenced by SNe can stay
in the inner region of the GC). 
SNe from FG stars formed later can suppress/truncate star formation,
if the SNe occur in the central region of the forming GC during SG star
formation there.
AGB ejecta that is influenced by SNe (i.e., ejecta located close to
SNe)  can not stay in the central region
of the forming GC. Accordingly, such ejecta is unlikely to
form new stars: little
self-enrichment in SG stars.
}
\label{Figure. 6}
\end{figure}

\begin{figure}
\psfig{file=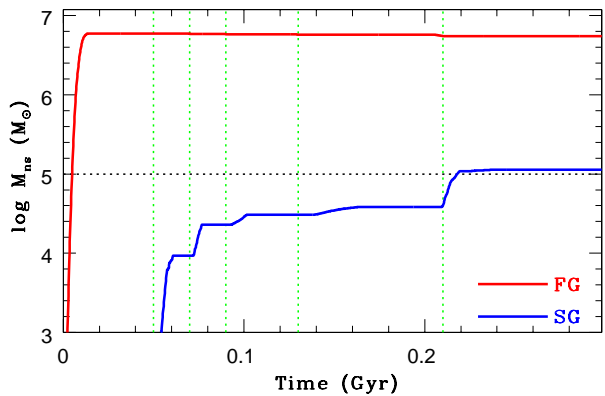,width=9.0cm}
\caption{
Time evolution of the total  mass of FG (red) and SG (blue) stars 
in the fiducial model. The epochs of gas ejection of AGB stars with different
masses are indicated by vertical green dotted lines, as shown in Fig. 1.
A horizontal black dotted line shows the total masses of SG stars observed
for typical old GCs of the Galaxy.
}
\label{Figure. 7}
\end{figure}

\begin{figure}
\psfig{file=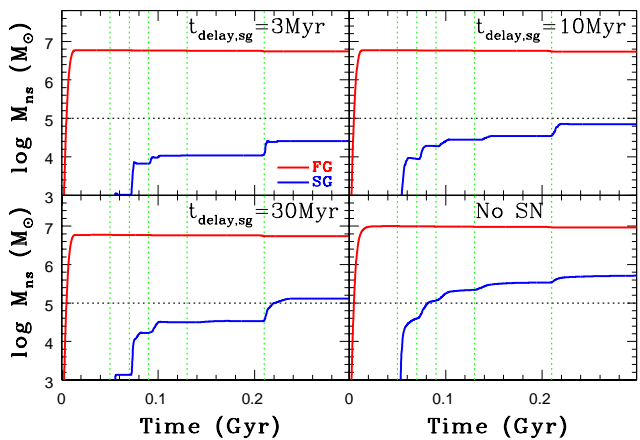,width=8.5cm}
\caption{
Time evolution of the total  mass of FG (red) and SG (blue) stars 
in the models with $t_{\rm delay, sg}=3$ Myr (upper right; M2),
$t_{\rm delay, sg}=10$ Myr (upper left; M3),
$t_{\rm delay, sg}=30$ Myr (lower left; M4), and
no SN feedback in both FG and SG formation (lower right; M5).
The epochs of gas ejection of AGB stars with different
masses are indicated by green dotted lines, as shown in Fig. 1.
}
\label{Figure. 8}
\end{figure}

Although a significant fraction of AGB ejecta can be converted into
new stars (SG),  about 70\% of the ejecta 
(in particle number) can not be converted into
new stars in the central region
of  the forming GC. One of the physical reasons for this
is that the ejecta can be  influenced by energetic SNe: if SN explosions
occur in the inner region of the forming GC , then the gas close to them
can be expelled from there. The other reasons is that the gas density
is not so high as the adopted threshold gas density for star formation.
Since discrete four epochs of SNe ($t_{\rm sn}$)
are assumed in the present simulation,
the influences of SNe on AGB ejecta could be under-estimated to some
extent. It would be possible that gas ejected from AGB stars
between interval of two discrete SN events can be accreted onto
the central regions of GCs. However, such gas accretion is less likely
because the differences of $t_{\rm sn}$ are typically small.

The derived timescale of SG formation that is much longer
than the lifetime of massive stars that explode
as massive SNe ($\sim [3-10]$ Myr) imply that the formation of such massive stars
need to be severely suppressed in SG star formation.
The FG stellar system grown through merging of hierarchical star cluster
complex in this model can finally have a very high mass 
($M_{\rm fg} = 5.47 \times 10^6 {\rm M}_{\odot}$).
It has a more diffuse distribution and   a large  effective radius
($R_{\rm e, fg} \sim 25$ pc at $T=113$ Myr) whereas
the SG stellar system has a very compact configuration 
with $R_{\rm e, sg}=1$ pc.
The mass ratio of SG to FG stars within  the central 1 pc ($=R_{\rm e, sg}$)
of the simulated GC at $T=113$ Myr is 1.9, which is consistent with
the observed fraction of GCs (C09).
The final total mass of the SG system is 
$M_{\rm sg}=1.13 \times 10^5 {\rm M}_{\odot}$, which is roughly similar to
the present typical total mass of SG stars (C09).
Therefore, the mass ratio of  SG to FG stars ($f_{\rm sg}$) is quite small
($\sim 0.021$), though most of FG stars form an outer stellar halo around the
simulated GC. 
The derived large $M_{\rm fg}$ means that the vast majority of FG stars
should be lost from the nested stellar systems, as discussed by
several authors already (e.g., D08, B11, Vesperini et al. 2010).
In the present fractal MC model, the initial MC mass should be 
quite large ($M_{\rm mc} \sim 10^7 {\rm M}_{\odot}$), which is 
even larger than the mass 
($\sim 3 \times 10^6 {\rm M}_{\odot}$)
of the most massive giant molecular cloud (GMC)
in the Galaxy (e.g., Solomon et al. 1987). 
This large required $M_{\rm mc}$ for GC formation within MCs might explain
why the Galaxy currently does not have GCs in formation.

Although it is not straightforward to estimate the timescale of
dynamical relaxation ($t_{\rm r}$) for the nested stellar system
in the fiducial model, we can derive $t_{\rm r}$ separately for
FG and SG stars using the formula by Spitzer \& Hart (1971).
The FG stellar system has
$t_{\rm r}=3.5 \times 10^{11}$ yr 
at $R=100$ pc
(for $M_{\rm fg}=5.47 \times 10^6 {\rm M}_{\odot}$),
which is much longer that the Hubble time.
The SG system has 
$t_{\rm r}=6.8 \times 10^{8}$ yr 
at $R=5$ pc
(for $M_{\rm fg}=1.13 \times 10^5 {\rm M}_{\odot}$).
The very long $t_{\rm r}$ of FG stars suggests that it is not possible for the 
entire FG and SG populations to be   mixed well within $\sim 10$ Gyr for the simulated 
GC, though the stars in the central region can be mixed together.
Thus, the central region of the GC can be dominated by SG stars for a long
timescale.

Fig. 5 shows that the simulated GC has an almost spherical distribution
of SG stars  in
the inner 2pc with a more elongated (elliptical) outer halo
of SG stars  for the $x$-$y$
and $x$-$z$ projection. 
The spherical distribution can be due to no rigid rotation of the 
MC in this model.
Fig. 6 demonstrates that there is a clear separation between the initial
bursty formation of FG stars and the later sporadic formation of SG stars
from gas of AGB stars with different masses.
This is due largely to a combination of (i) efficient
removal of remaining gas by SN explosion and (ii) long $t_{\rm agb}$
of AGB stars ($> 4 \times 10^7$ yr).
In the present model, continuous gas ejection from AGB
stars with different masses can not be properly modeled owing to
the strong limitation of gas particles numbers adopted in the simulations.
Therefore,  there are five peaks in the ejection rate 
($dM_{\rm ej}/dt$) of gas from
AGB stars, which correspond to the commencement of AGB phases of
intermediate-mass stars with 
$m_{\rm s}=[7-8] {\rm M}_{\odot}$,
$[6-7] {\rm M}_{\odot}$,
$[5-6] {\rm M}_{\odot}$,
$[4-5] {\rm M}_{\odot}$,
and $[3-4] {\rm M}_{\odot}$, respectively.
Soon after each epoch of gas ejection from AGB stars,  the star formation rate
(SFR) for SG stars can significantly increase owing to the increased
mass density of gas in the central region of the proto-GC.
This apparently sporadic increase in SFR results simply from
the adopted model for gas ejection from AGB stars in the present study:
this should not be interpreted as the formation of discrete sub-populations
within a single GC.

Fig. 6 shows that SN explosion is 
ongoing even when AGB stars are ejecting gas in
the forming GC ($T>50$ Myr). Since SN explosion is not assumed to occur
in stars formed from gas ejected from AGB stars (i.e.,
SG stars)  in the fiducial model,
all SNe in Fig. 6 are from FG stars.
These FG stars are formed from (original) gas particles that were
not influenced by SNe owing to its location  being not close enough to SNe
at earlier times ($T<40$ Myr). If original gas particles are  influenced by
SNe, then they gain energy and momentum and are chemically polluted
by metals of the SNe. Accordingly,  such gas particles can not settle down to
the central region of the forming GC.
Thus, gas particles that are later ($T>50$ Myr)
converted into FG stars are much less
chemically enriched by SNe.
The time lag between gas ejection of AGB stars and SG star formation
is rather short ($<3$ Myr), because accretion of AGB ejecta onto the 
inner region of the forming GC can rapidly proceeds
after almost all massive
stars explode.

The formation of SG stars from AGB ejecta is possible, only if
the ejecta is not located to the vicinity of a SN (i.e., only if
it does not gain energy and momentum from the SN).
AGB ejecta can be expelled from the forming GC if it is influenced by
SN feedback effects. Accordingly,  SG star formation at $T>50$ Myr in Fig. 6
is from AGB gas particles that are not influenced by SNe:
The apparent coincidence of SN explosion and onset of AGB phase does not
mean chemical pollution of AGB ejecta by SNe.  
It should be noted here that these later  SNe are
not necessarily
located in the central region of the forming GC. 
SNe of FG stars formed later in the central region of the forming GC
can expel the remaining AGB ejecta so that SG star formation can be
severely suppressed or temporarily truncated.

A very small fraction of cold gas can not be completely 
ejected from the MC and therefore used for further star formation
after its accretion onto the stellar system developing in the MC.
This gas is not chemically polluted by SNe, because it is initially
located in the outer part of the gas cloud (or because it is not
located close to SNe).
Given the pristine nature of the gas, 
AGB ejecta accumulated in the stellar system can be mixed with
(or `diluted by')  the gas, though such dilution is not so efficient
owing to the small mass of the gas.
This dilution of AGB ejecta by pristine gas is one of essential ingredients
of chemical evolution models for GCs with multiple stellar populations
(e.g. Fenner et al. 2004; Bekki et al. 2007; D10),
and the origin of such pristine gas has been discussed in 
previous works (e.g., D08, D10, B11, B17). 
The present study accordingly suggests that original cold gas that is not
influenced by SN explosion can be used for dilution
of AGB ejecta in forming GCs.
However, the amount of such cold gas in the present study is too small
(an order of $ \sim  10^4 {\rm M}_{\odot}$) within the central 10pc of the proto-GC
in comparison with the required one in previous models.

\begin{figure}
\psfig{file=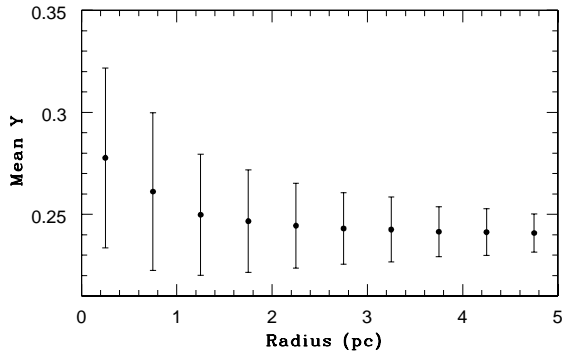,width=8.0cm}
\caption{
The projected radial profile of mean helium abundances ($Y$) of GC stars
(FG+SG) at $T=112$ Myr in the fiducial model.
An error bar indicates the dispersion of $Y$ in each bin.
}
\label{Figure. 9}
\end{figure}

\begin{figure}
\psfig{file=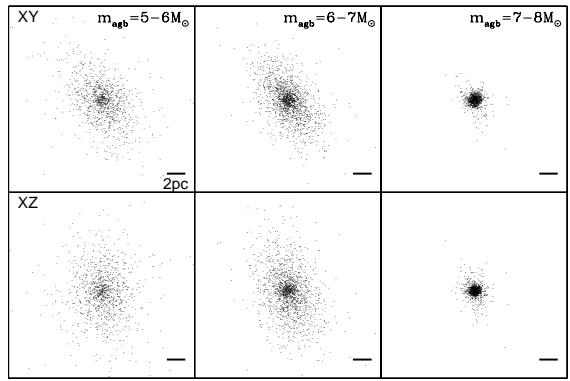,width=8.0cm}
\caption{
The same as Fig. 1 but for SG stars formed from gaseous ejecta
of AGB stars with masses of $[5-6] {\rm M}_{\odot}$ (left),
$[6-7] {\rm M}_{\odot}$ (middle),
and  $[7-8] {\rm M}_{\odot}$ (right),
}
\label{Figure. 10}
\end{figure}

Fig. 7 demonstrates that $M_{\rm sg}$ can exceed $10^5 {\rm M}_{\odot}$
only after gaseous ejecta of low-mass AGB stars with 
$m_{\rm agb}=[3-4] {\rm M}_{\odot}$ is accumulated onto the 
central region of the proto-GC and converted into new stars.
This prolonged SG formation over  $\sim 0.2$ Gyr in the MC
can constrain the IMF of SG stars, as discussed later in this paper.
The SFE of SG stars  ($\epsilon_{\rm sf, sg}$)
is  0.3 in this fiducial model, which suggests that the mass budget
problem is even severer than ever thought in previous models
in which $\epsilon_{\rm sf, sg}$ is assumed to be rather high.
These results suggest that the threshold MC mass ($M_{\rm mc, th}$) beyond which
typical GCs with $M_{\rm sg} \sim 10^5 {\rm M}_{\odot}$ can be
quite large:
\begin{equation}
M_{\rm mc, th} \approx 10^7 {\rm M}_{\odot}.
\end{equation}
It should stressed here that the original $M_{\rm sg}$ in GCs
can be significantly larger than 
$10^5 {\rm M}_{\odot}$, because GCs could have lost SG stars from
tidal stripping and long-term internal dynamical processes driven by two-body
dynamical relaxation. Thus, $M_{\rm mc, th}$ can be larger than
the above value.

The SFE of FG stars in this model is high ($\sim 0.5$), which
is likely to be over-estimated, because
the present study does not include suppression of star formation
from original gas through ionizing photons and stellar winds of massive
stars. These suppression effects
are properly modeled in recent simulations of MC evolution
(e.g., Dale et al. 2014). Since  $\epsilon_{\rm sf, fg}$ can control
the total mass of AGB ejecta from which SG stars can be formed,
the present study without gas ionization and stellar winds
of massive stars (before SNe) could also  overestimate the total mass
of SG stars in each simulated GC.

Fig. 8 shows that $M_{\rm sg}$ is smaller than
$10^5 {\rm M}_{\odot}$ required for the formation of genuine GCs
in the models (M2 and M3)  with SNe from SG stars,
if $t_{\rm delay, sg} \le 100$ Myr.
The model M2 with $t_{\rm delay, sg}=3$ Myr shows 
$M_{\rm sf}=2.6 \times 10^4 {\rm M}_{\odot}$,
which implies  that $M_{\rm mc, th}$ should be even significantly
higher than $10^7 {\rm M}_{\odot}$ derived for M1.
This result  means that SNe from SG stars themselves can severely suppress
the efficient conversion of AGB ejecta into new stars, because
energetic SN feedback effects brow away the ejecta from the central
region of the proto-GC. This furthermore suggests that the formation of
high-mass SNe with short lifetimes ($ \le 100$ Myr) from
SG stars need to be  
truncated for the formation of GCs with 
$M_{\rm sg} > 10^5 {\rm M}_{\odot}$.
This results implies that the upper-mass cutoff of the IMF for SG stars
needs to be quite low ($<10 {\rm M}_{\odot}$) for the efficient 
formation of SG stars from AGB ejecta ($\epsilon_{\rm sf,sg} \sim 0.3$).
The simulated GC can have 
$M_{\rm sg} > 10^5 {\rm M}_{\odot}$ in the early phase of GC formation
($T<0.1$ Gyr) in the model (M5) without SNe from FG and SG stars, because
AGB ejecta can be quickly accreted onto the central region of the GC without
being influenced by SN feedback effects.

As shown in Fig. 9,
the simulated GC in the fiducial model
has a negative radial gradient of helium abundance ($Y$) 
within the central 3 pc of the GC. The higher $Y$ in the inner region
results from the higher mass fraction of SG stars with higher $Y$
due to their more compact distribution.
The dispersion in $Y$ is also higher in the central region, which reflects
the fact that new stars can be formed from gaseous ejecta from AGB
stars with different masses and thus different $Y$.
The outer part of the simulated GC is dominated by FG stars with $Y=0.24$
so that the dispersion in $Y$ can be rather small.
These negative $Y$ gradient and higher dispersion in $Y$ in the inner
region of the simulated GC can be seen in almost  all massive
MC  models of the present
study.
The adopted simulation code does not allow us to discuss the long term
($>10^9$ yr) dynamical evolution of a GC through two-body relaxation
processes. Accordingly, it remain unclear whether the derived 
negative radial gradients of
$Y$ in the simulated GCs can persist for $10^{10}$ yr.
It could be possible that
such initial $Y$ gradients can be kept as they are 
in massive GCs with long relaxation timescale, such as
$\omega$ Cen.

In the present model of GC formation,
gaseous ejecta from more massive AGB stars
can be accumulated onto a diffuse
FG stellar system  earlier so that it can be converted into
a very compact stellar system. 
Most gaseous ejecta from less massive AGB stars
can not reach the very central region of the forming GC,
instead, it can be accreted later onto the surrounding region of the compact SG stellar
system formed earlier (the ejecta forms a disky structure).
As a result of this,  SG stars formed from
low-mass AGB stars have more diffuse spatial distributions.
Fig. 10 clearly demonstrates that SG stars formed from massive AGB stars
with masses of $[7-8] {\rm M}_{\odot}$ have a  more compact distribution. 
If more massive AGB stars with $[7-8] {\rm M}_{\odot}$
 can eject gas with higher $Y$
than less massive ones with $[5-6] {\rm M}_{\odot}$,
then this result in Fig. 10
implies that there can be differences in the spatial
distributions between SG stars with different $Y$.
The chemical yield model of AGB
stars  from K10 adopt in the present study
predicts such a trend of increasing $Y$ with increase AGB star mass.

\begin{figure*}
\psfig{file=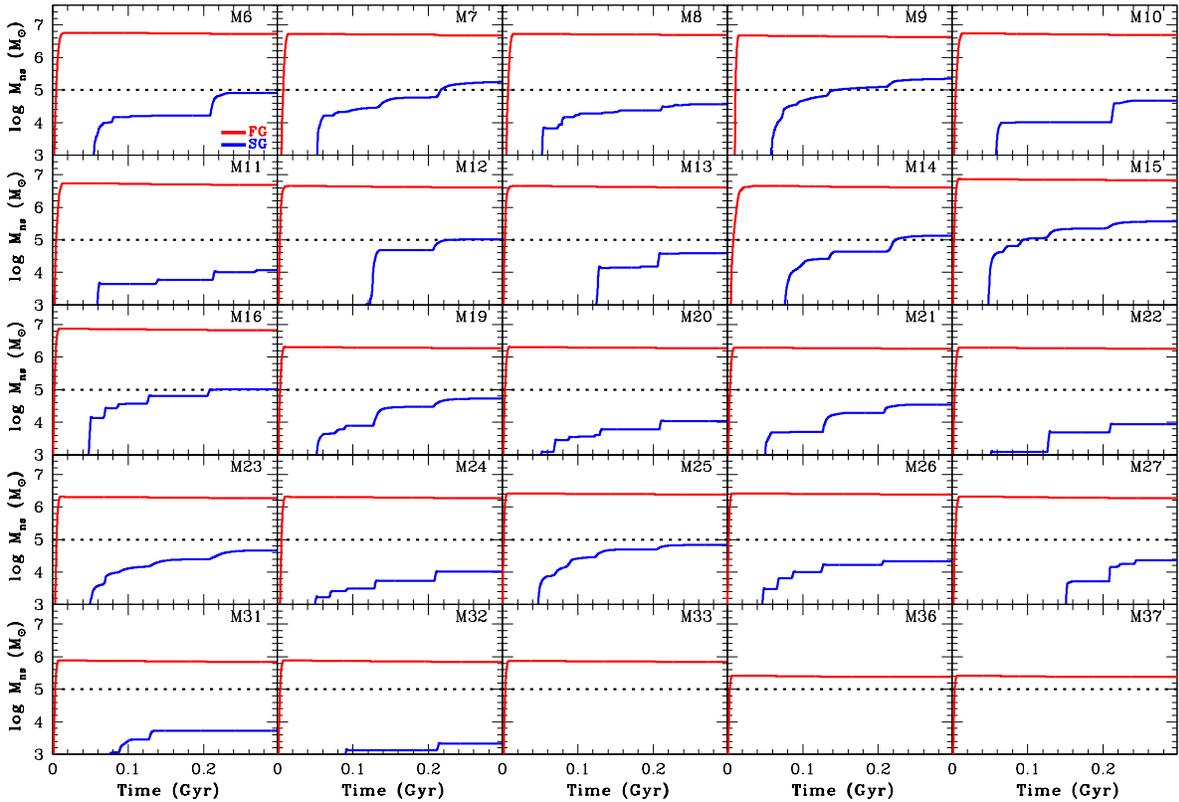,width=16.0cm}
\caption{
The same as Fig. 7 but for 25 representative models in the present study.
The model ID is given in the upper right corner of each panel. 
25 models are selected from 38 investigated in the present study.
}
\label{Figure. 11}
\end{figure*}

\subsection{Parameter dependence}

Formation processes of GCs (i.e., `two-stage' FG and SG star formation)
and physical roles of feedback effects of SNe and AGB stars in
star formation histories (SFHs) of GCs are essential the same between
different models. However, the details of the two-stage GC formation
processes and SFHs depend on model parameters. Furthermore,
$M_{\rm sg}$ in some models 
with lower $M_{\rm mc}$ is too small for the simulated GCs to be
identified as GCs with multiple stellar populations. Such GCs dominated
by FG stars are regarded as `failed GCs' in the present study
and might be better labeled as low-mass SCs.
The time evolution of $M_{\rm fg}$ and $M_{\rm sg}$
and final $M_{\rm sg}$ and $\epsilon_{\rm sf, sg}$ 
for representative models in the present study are summarized  
in Figs. 11 and 12, respectively.
The dependences of the present results on the model parameters
are summarized as follows. \\

(1) As shown in Fig. 11, only 
some of the very massive MC models with $M_{\rm mc} = 10^7 {\rm M}_{\odot}$
can show $M_{\rm sg} \approx 10^5 {\rm M}_{\odot}$, which is required
for the present-day typical GCs. This result suggests
that there is a threshold MC mass beyond ($M_{\rm mc, th}$)
which GCs can be formed
from MCs. Even the simulated GCs in the models
with $M_{\rm mc} = 3 \times 10^6 {\rm M}_{\odot}$ can not have
SG stellar systems with $M_{\rm sg} \approx 10^5 {\rm M}_{\odot}$. 
This means that even the most massive Galactic MCs 
with $M_{\rm mc} = 3 \times 10^6 {\rm M}_{\odot}$ (e.g., Solomon et al. 1987)
are unlikely to form GCs. Thus the possible presence of $M_{\rm mc, th}$
is a physical reason why the Galaxy is currently not forming massive GCs.
\\ 

(2) The simulated SCs (or `failed GCs') in the models with 
$M_{\rm mc}= [1-3] \times 10^6 {\rm M}_{\odot}$ (M19 - M35)
can contain SG stars, however, both $f_{\rm sg}$ and $\epsilon_{\rm sf, sg}$
are significantly
lower than those derived for  the models with
$M_{\rm mc}=  10^7 {\rm M}_{\odot}$.
These results imply that $f_{\rm sg}$ can be quite diverse, however,
the observed $f_{\rm sg}$ is not so diverse (C09).
Although the results only for three models are shown 
in the present study (Table 1 and Fig. 11),
it is confirmed that
simulated SCs in the models with 
$M_{\rm mc}=  3 \times 10^5 {\rm M}_{\odot}$ (M36 - M38) do not contain SG stars
at all for models 
with different parameters.
Such an inability of SG star formation in low-mass MC results from
the fact that AGB ejecta can not form high-density gaseous regions
within stellar systems composed of FG stars. A significant fraction AGB ejecta
with a wind velocity of 10 km s$^{-1}$
can escape from the low-mass MCs with shallower gravitational potentials.
The final SCs in these models can not be regarded as genuine GCs, 
and they are more similar to open clusters with single stellar populations. \\
\\

(3) Compact stellar systems with $M_{\rm sg} \sim 10^5 {\rm M}_{\odot}$
can be formed within MCs with $M_{\rm mc} = 10^7 {\rm M}_{\odot}$,
regardless of whether they have initial rotation (M6). However,
$\epsilon_{\rm sf,fg}$ and $\epsilon_{\rm sf,sg}$ in the rotating MC model  M6 are slightly
smaller than those in M1 without rotation. This less efficient SF in FG and SG
stars in rotating MCs can be seen in models with different $M_{\rm mc}$ 
(e.g., M21 and M22). This suggests that initial rotation of GC-hosting MCs can also
control $M_{\rm fg}$ and $M_{\rm sg}$ and thus the present-day masses of GCs. 
It should be also noted here that
binary GCs can be formed in the models with
$f_{\rm rot} \ge 0.1$  in some low-mass MC models. 
This binary SC formation in fractal MCs will be discussed in our forthcoming papers.
\\

(4) Regardless of $M_{\rm mc}$,
the two-stage GC formation process does not
depend strongly on $D_3$ (e.g., M7 vs M8 for $M_{\rm mc}=10^7 {\rm M}_{\odot}$). 
The suppression of SG star formation by SNe from SG stars and  the resultant lower
$M_{\rm sg}$ can be clearly seen in the models with $D_3=2.4$
for $M_{\rm mc}=10^7 {\rm M}_{\odot}$  (M8 vs M9),
$M_{\rm mc}=3 \times 10^6 {\rm M}_{\odot}$  (M23 vs M24),
and $M_{\rm mc}=10^6 {\rm M}_{\odot}$  (M33 vs M34),
Here the results in the models with  $D_3=2.4$ are not described for
$M_{\rm mc}=3 \times 10^5 {\rm M}_{\odot}$ ,
simply because they do not show any SG  star formation.
These results for $D_3=2.4$  combined with those for
$D_3$ demonstrate that SN feedback effects are the most important
physical effect for SG formation.
These also imply that the IMF for the SG star formation should be top-light
(almost no massive SNe) for GCs to have  significant fractions of SG stars.
This point is discussed later in this paper. \\

(5) Threshold densities for star formation ($\rho_{\rm th}$)
can also control the time evolution of $M_{\rm fg}$ and $M_{\rm sg}$
in the sense that  $M_{\rm fg}$ and $M_{\rm sg}$ can be smaller
for larger $\rho_{\rm th}$ for models with different $M_{\rm mc}$.
The final $M_{\rm sg}$ can be significantly smaller in the models
with $\rho_{\rm th}=10^5$ than in those with $\rho_{\rm th}=10^4$ (i.e.,
strong suppression of SG formation).
Although this is initially expected for the adopted star formation model,
this has some implications on GC formation, which is discussed later in this paper.
MCs with higher initial mass densities can show larger $M_{\rm sg}$
(e.g.. M14 vs M15). The high-density massive MC model M16 with SNe from
SG stars 
show $\epsilon_{\rm sf, sg}=0.24$ and $M_{\rm sg}=1.0 \times 10^5 {\rm M}_{\odot}$.
This result implies that if massive MCs have rather high initial densities,
then SNe from SG stars can not so strongly suppress SG star formation.
\\

(6) The strong tidal field of a MC-host dwarf galaxy does not so strongly influence
the formation processes of GCs. Compact SG systems can be formed from
AGB ejecta in the models with galactic tidal fields (M17, M18, M29, and M30).
Final $M_{\rm sg}$ and $\epsilon_{\rm sg}$ in the models with galactic tidal
fields are appreciably smaller than those without. For example,
$\epsilon_{\rm sf, sg}$ is 0.12 for $r_{\rm p}=0.3$ kpc (M17)
0.21 for $r_{\rm p}=1$ kpc (M18).
Galactic tidal fields can be important for disintegrating the more diffuse
FG stellar systems, as suggested by previous numerical simulations of
GC formation (D08 and B11). The present models with and without
galactic tidal fields show a correlation between $M_{\rm sg}$
and $\epsilon_{\rm sf, sg}$
(See Fig. 12).
This is due partly to  SFE of SG stars being  higher in MCs
with higher masses. \\

(7) As shown in Fig. 13 for the fiducial model,
GC stars do not show large metallicity spreads 
in the simulated massive Gs. SG stars show a smaller
metallicity spread than FG stars,  because they can be formed
from almost pure AGB ejecta without significant chemical pollution
by SNe. However,
$\Delta$[Z/H] can be as large as 0.04 dex for FG stars. Fig. 13 shows
that $\Delta$[Z/H] of GC stars (FG+SG) within 20 pc
is slightly larger than the observed
$\Delta$[Z/H] ($<[0.02-0.03]$ dex) for 17 Galactic GCs (Caretta et al. 2010).
Gas remaining in the outer part
of the the GC ($R>20$ pc) shows a large metallicity spread, because
the gas was expelled by SNe after being chemically polluted by SNe.
It is not observationally clear whether FG stars show a larger $\Delta$[Z/H]
than SG stars in a GC, as predicted in the present study.
If FG stars in observations do not show larger $\Delta$[Z/H],
then the present model would need to be revised in terms of chemical pollution
of original gas by SNe.

(8)
In the present parameter study, mixing of AGB ejecta with SN ejecta is not 
so well resolved in all models,
 because the minimum number of AGB particles (or
any gas particles) around one SN is set to be 8. This could cause
an under-estimation of SN effects on AGB ejecta. For example,
the present study could have over-estimated $M_{\rm sg}$ 
and under-estimated the metallicity spreads among SG stars owing
to the possible under-estimation of the mixing of SN and AGB ejecta.
This issue will need to be discussed in our future simulations with a
much better spatial resolution for the interaction between SN and AGB
ejecta. The potential problem of AGB ejecta being contaminated by
hot gas from SNe has not been convincingly 
(and fully) solved in the present study with a spatial resolution of an order
of 0.1pc.

\section{Discussion}

\subsection{Necessity of top-light  IMF in SG star formation
in self-enrichment scenarios}

Most old globular clusters (GCs) in the Galaxy to have star-to-star
abundance spreads in light elements (e.g., C, N, and O)
and the observed Na-O anti-correlation has been
considered to be one of the characteristic features of GCs (e.g., C09).
If the majority (70\%) of stars 
with enhanced Na and depleted O (i.e., SG stars)
are formed from 
gas ejected from FG stars with normal Na and O,
then the original total mass of FG stars in a GC can be inferred from
the present-day total mass of SG stars ($M_{\rm sg}$). Previous studies
suggested that $M_{\rm fg}$ should be much more massive
than $M_{\rm sg}$ (i.e., the mass budget problem).
In the following,  we discuss this mass budget problem
in the context of self-enrichment scenarios based on SG formation
from AGB ejecta.

The mass budget problem can be formulated as follows (B17):
\begin{equation}
M_{\rm fg}=1.4 \times 10^7
{ ( \frac{ \epsilon_{\rm sf, sg} }{0.1} ) }^{-1}
{ ( \frac{ f_{\rm ej,fg} }{0.1} ) }^{-1}
( \frac{ M_{\rm sg,0} }{1.4 \times 10^5 {\rm M}_{\odot} } )
{\rm M}_{\odot},
\end{equation}
where 
$f_{\rm ej, fg}$ is the mass fraction of gas ejected from AGB stars
of FG and
the star formation efficiency in SG star formation
($\epsilon_{\rm sf, sg}$) is assumed to be 0.1, as shown in the models with
SN feedback effects on SG star formation.
As long as a canonical IMF is assumed,
such a small $f_{\rm ej, fg}$ and SN feedback effects on SG formation
(i.e., low $\epsilon_{\rm sf, sg}$)
are inevitable outcomes.
The derived  $M_{\rm fg}$ is underestimated to some extent, because 
mass loss of SG AGB stars are not considered.
Nevertheless $M_{\rm fg}$ appears to be too large,
which means that 99\% of FG stars needs to be lost to form
GCs with the observed mass fractions of FG and SG stars.
Thus, if the IMF of SG stars is a canonical one, then
the mass budget problem is much more severe than ever thought.

The mass budget problem can be less severe, if the IMF of FG stars
is top-heavy (i.e., a larger fraction of AGB stars thus a higher $f_{\rm ej}$)
and if the IMF of SG stars are top-light (i.e., a smaller fraction of SNe).
Probably, the formation of massive SNe with shorter lifetimes 
need to be completely shut down to alleviate the mass budget problem.
Such suppression of massive
star formation  was already pointed out by D08, though they did not discuss this
in a quantitative manner.
A key question in any self-enrichment scenarios is therefore whether or not
the formation of massive stars ($m_{\rm s} \ge [30-120] {\rm M}_{\odot}$) 
can be really severely suppressed (i.e., `top-light' IMF)
in SG star formation. 
Although our previous simulations of SG formation in dense stellar systems
of FG stars investigated star formation processes (B10, B11),
the IMF of SG stars could not be investigated owing to the adopted resolution.
It is accordingly our future theoretical  study
to investigate the slope and the upper-mass cutoff of the IMF for SG stars 
using numerical simulations that resolve star-forming cores.
If our future study on this issue reveals that there is no plausible
theoretical reason for the top-light IMF of SG stars,
then we would need to discard self-enrichment scenarios as a major mechanism
of GC formation.

\begin{figure}
\psfig{file=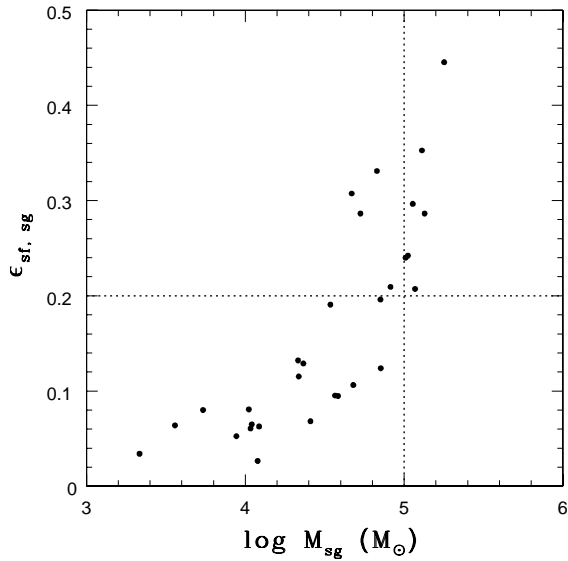,width=8.0cm}
\caption{
Star formation efficiencies (SFEs) of SG stars ($\epsilon_{\rm sg}$)
as a function of $M_{\rm sg}$ for all models investigated
in the present study.
The vertical and horizontal dotted  lines represent the observed typical mass
of SG stars in the Galactic GCs and the threshold SFE ($0.2$)
above which bound clusters can be formed from gas clouds 
(e.g., Hills 1980; Elmegreen \& Efremov 1997).
}
\label{Figure. 12}
\end{figure}

\subsection{If self-enrichment scenarios are not viable,
then why can not secondary star formation occur  ?}

If self-enrichment scenarios based on gaseous ejecta from AGB stars
are not viable for GC formation with
multiple stellar populations,
then it needs to be clarified why secondary star formation can not
occur in forming GCs with a plenty of AGB ejecta.
 As shown in previous
theoretical works on the formation of massive  GCs,
gaseous ejecta of AGB stars can be well retained in the central regions of the
GCs (D08 and B11). Therefore, some physical mechanisms need
to operate so as to
suppress conversion from gas into new stars in dense stellar systems. 
Recent galaxy-scale numerical simulations of star formation histories in
luminous and dwarf disk galaxies have demonstrated that star formation
can be severely suppressed by photo-electric heating (PEH) of cold gas by dust
(Bekki 2015; Forbes et al. 2016). The gas accumulated onto FG stellar systems
from AGB stars can be dust-rich so that PEH effects can be strong if 
there is an enough
amount of stellar radiation from FG stars in the central regions of proto-GCs.
However, if AGB ejecta is diluted by pristine metal-poor gas, which is required
for chemical evolution model of GCs,
then such PEH effect could be weak.  
Accordingly, we need to investigate how such PEH effect can influence
the secondary star formation processes in proto-GCs.

It would be also possible that high number densities of stars in proto-GCs
can totally prevent gravitational instability of gas that leads to star formation.
The spatial resolution ($\sim 0.4$ pc) of the present simulation is not good enough to
investigate the very small-scale ($<<1$ pc) formation processes
of each individual stars. If dense stellar environments can really prevent
secondary star formation, then it needs to be understood how and where
AGB ejecta can be lost in proto-GCs.
Ram pressure stripping of AGB ejecta by warm and hot ISM 
of GC-host galaxies  could be a candidate mechanism for the removal of AGB ejecta,
if proto-GC can pass through such ISM. It might be also possible that
GCs can lose their AGB ejecta by ram pressure
when they orbit around the halos of their host
luminous galaxies like the Galaxy.
If self-enrichment scenarios need to be discarded,
a crucial question is when GCs achieved star-to-star internal
abundance spreads during their formation histories.
One idea is that new stars of proto-GCs already had 
star-to-star internal abundance spreads
before their first SNe explode.
It is not clear, however, how such chemical enrichment can proceed
within $\sim$ 3 Myr (before first massive SNe explode).

It should be noted here that the above discussion is based on
secondary star formation of gas ejected from AGB stars. An alternative
self-enrichment scenario based on gaseous ejecta from fast-rotating
massive stars (FRMS)  has been already discussed by several authors
(e.g., Decressin et al. 2007). This FRMS self-enrichment scenario
has no problem associate with later SNe that can severely suppress
secondary star formation. However, it is not clear in this FRMS scenario
how SG star formation can be completed well before SN explosions of FG stars,
which can expel all of the remaining gas within GC-forming molecular clouds.
We here do not discuss how to avoid this potentially serious problem,
because it is beyond the scope of this paper to investigate
the FRMS scenario in detail. in detail.

\begin{figure}
\psfig{file=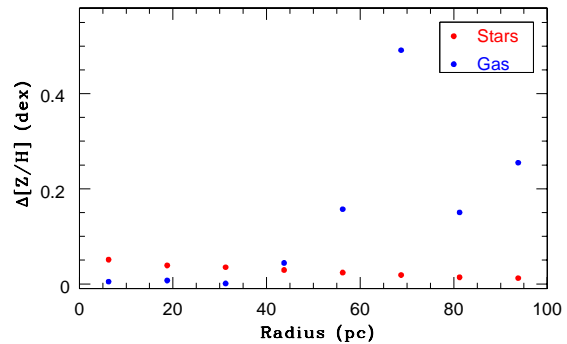,width=8.0cm}
\caption{
Internal metallicity spreads ($\Delta$[Z/H]) for GC stars (red) and 
gas (blue) as a function of radius ($R$ from the GC center)
in the simulated massive GC for the fiducial model.
The internal [Z/H]  spread of gas is  smaller than that of
the stars for $R<20$ pc, 
because only the gas that is not
chemically polluted (not expelled by SNe)
can remain in the central region of the GC.
The [Z/H] spread
for SG stars formed from such gas (less chemically contaminated by SNe)
can be smaller than that of FG stars.
The [Z/H] spread of gas in the outer part of GC is significantly larger
than that of the stars ($R>60$ pc), because the gas is the AGB ejecta
expelled by SN explosions  after being chemically polluted by SNe.
The [Z/H] spread of GC  stars is slightly larger than the observed
small spread ($<[0.02-0.03]$ dex)  by Carretta et al. (2010).
}
\label{Figure. 13}
\end{figure}

\subsection{Metallicity spreads in GCs}

Although Lee et al. (2009) investigated the color magnitude
diagrams ($V$ vs $b-y$ or $hk$) of the Galactic GCs
and found evidence of Ca abundance spreads in the 7 GCs,
Carretta et al. (2010) showed that Ca abundance spreads among FG and SG stars
in 17 Galactic GCs are less than $0.02-0.03$ dex.
The most massive
Galactic GC $\omega$ Cen and 8 GCs (e.g., M22) have been observed
to  show [Fe/H] spreads among
GC stars so far (e.g., Freeman  \& Rodgers 1975; Marino et al. 2015).
The apparent lack of internal  [Fe/H] spreads among GC stars 
in most  GCs implies that self-enrichment by SNe did not proceed efficiently
at GC formation for some physical reasons. Since these GCs show
anti-correlations between light elements, which could be due to self-enrichment
by AGB stars,  the physical mechanisms that suppress self-enrichment
by SNe in GC formation need to be understood clearly.
Nakasato et al. (2000) investigated star formation histories
of proto-GC clouds with the masses of $10^6 {\rm M}_{\odot}$,
sizes of $150-300$ pc, and initial temperature of $10^4$ K using
their original SPH simulations with feedback effects
chemical enrichment by SNe.
They found that although star formation
in shell-like gaseous structures formed
through compression of gas through SNe feedback effects is possible,
self-enrichment is not seen to occur in all of their models. 

The present study have demonstrated that 
although new star formation from gas contaminated
by SNe is possible, the mass fraction of such stars is quite small.
Furthermore, most of such stars can be formed mostly in
the shocked gas that are distant from the main GC-forming regions,
and therefore they can
not be finally within the central
regions of  GCs. 
As a result of this, the mass fraction of such (SG) stars with [Fe/H] 
by more than 0.05 dex
larger than (FG) stars formed from original cold gas of GC-forming GCs
is quite small within the central 10pc of GCs.
This implies that the apparent lack of [Fe/H] spreads in typical
GCs is due largely to SN feedback effects in GC-forming MCs.
Baumghardt et al. (2008) shows that the stellar masses of GCs
required for self-enrichment 
by SNe is more than $10^7 {\rm M}_{\odot}$, which means
that original GC-hosting MC should be very massive
($ \sim 10^8 {\rm M}_{\odot}$) for a reasonable star formation efficiency ($\sim 0.1$).
Therefore, typical GCs are unlikely to have [Fe/H] spreads.

The most massive Galactic GC $\omega$ Cen has been suggested to originate from
a nucleated dwarf galaxy (e.g., Freeman 1993; Bekki \& Freeman 2002),
where its deep potential well could retain ejecta from SNe for further
star formation. Although other eight 
`anomalous' GCs with [Fe/H] spreads could be also
from defunct nucleated dwarfs like $\omega$ Cen,
it remains unclear what physical mechanisms
are responsible  for their [Fe/H] spreads.
One of intriguing observational results is that
some of anomalous GCs with metallicity spreads (e.g., M22)  also show
abundance spreads in $s$-process elements (Marino et al. 2011).
The observed spreads in $s$-process elements could be due to
star formation from gas polluted by  AGB stars.
The present study has shown that SG stars have the same metallicities
as those of FG stars, because SG stars can be formed from AGB ejecta 
only after SN explosion expel the remaining gas of MCs.
Therefore it appears unlikely that simple self-enrichment
scenarios can explain the origin of abundance spreads both in
[Fe/H] and $s$-process elements.

Bekki \& Tsujimoto (2016) have recently demonstrated that 
merging between massive GCs with initially different [Fe/H]
in their host dwarf galaxy  is possible,
which ends up with a bimodal [Fe/H] distribution that is
consistent with observations for M22 (Marino et al. 2011).
They suggested that other anomalous GCs could be also  formed from
GC merging within dwarf galaxies with different star formation
histories.
It could be possible that only very massive GCs like
$\omega$ Cen experienced self-enrichment by SNe at their formation:
[Fe/H] spreads in GCs alone do not necessarily mean star formation
from gas polluted by SNe within their host MCs.
Although GC merging is a promising mechanism for the origin
of anomalous GCs, it has not reproduced several chemical abundances
of their stars in a fully self-consistently manner (e.g., abundance spreads
in C+N+O). Thus,
there are still puzzling observational results on these GCs,
which need to be addressed in our future papers.

\section{Conclusion}

We have investigated the formation processes of GCs with
multiple stellar populations 
within massive molecular clouds (MCs) with fractal structures using our original
hydrodynamical simulations with star formation,
feedback effects of SNe and AGB stars, and chemical enrichment
by these stars. The key parameters of the simulations are 
the masses ($M_{\rm mc}$), sizes ($R_{\rm mc}$), 
ratios of rotational energy to total kinetic energy ($f_{\rm rot}$)
and fractal dimensions ($D_3$)
of GC-forming massive MCs.
We have analyzed the physical properties of new stars
formed from original pristine gas (first generation of stars; `FG')
and from gaseous ejecta of AGB stars (second generation; `SG').
We have also investigated (i) the models with and without
SN feedback effects in the formation of  SG stars
(ii) those in which
tidal field of dwarf galaxies hosting GCs are included.
The principal results are as follows: \\

(1) Bound massive clusters  of FG stars 
can be first formed from merging of hierarchical star cluster (SC) complexes that
are developed from fractal gaseous structures of massive cold  MCs
with $M_{\rm mc}=10^7 {\rm M}_{\odot}$.
During merging of low-mass SCs within MCs,
gas ejected from SNe can interact with the surrounding
pristine gas of GC-forming MCs. SNe of very massive stars
with $m_{\rm s}=[60-120] {\rm M}_{\odot}$ can brow off cold gas
from the very early stage of GC formation.
After almost all of the cold gas is expelled from proto-GCs by
SNe with different masses,
gas ejected from AGB stars can be accumulated into the central
regions of proto-GCs, where the AGB ejecta is converted into SG stars.
This formation process of SG stars from AGB ejecta
is consistent with the results of our previous 3D hydrodynamical  simulations
of GC formation 
(B10, B11). \\


(2) Most SG stars can be formed from gaseous ejecta in
the central regions of FG stellar systems for all massive MC models.
The spatial distributions of SG stars are therefore initially  more 
compact than those of FG stars in GCs for all models
with different parameters.  
Since SG stars can be formed from ejecta of AGB stars with
different masses,
there can be significant differences in chemical abundances between SG stars. 
GC stars (FG and SG)  show negative gradients of helium ($Y$) abundances
(i.e., higher $Y$ in the inner regions)
within the central 3 pc of the simulated GCs.
SG stars formed from high-mass AGB stars ($m_{\rm s}=7-8 {\rm M}_{\odot}$)
are more centrally concentrated than those from low-mass ones
($m_{\rm s}=4-5 {\rm M}_{\odot}$). This suggests that
there is significant differences in spatial distributions
between SG stars with different helium abundance ($Y$), 
because high-mass AGB stars
can eject gas with higher $Y$. \\

(3) There is a threshold MC mass ($M_{\rm mc,th}$) beyond which
$M_{\rm sg}$ can be as large as the observed value of typical
GCs ($\sim  10^5 {\rm M}_{\odot}$). This $M_{\rm mc,th}$
is as large as $10^7 {\rm M}_{\odot}$ for a canonical
(Salpeter) IMF and it can be smaller for more top-heavy IMFs.
The final masses of 
FG stars
can be quite large ($M_{\rm fg} \sim 5 \times  10^6 {\rm M}_{\odot}$) 
in the simulated GCs for $M_{\rm mc}=10^7 {\rm M}_{\odot}$
and the large fraction of FG stars reside  in the halo regions of the proto-GCs.
This  means that the vast majority of the FG stars need to be lost for
the simulated proto-GCs to become genuine GCs dominated by SG stars. 
This required removal of FG stars has been extensively discussed by
previous simulations already. \\

(4) The two-stage GC formation process (i.e., SG formation
after FG formation)  through
merging of hierarchical SC complexes  does not depend strongly on
$M_{\rm mc}$, $R_{\rm mc}$, $f_{\rm rot}$, $D_3$,
and tidal fields of MC-host dwarfs, though the physical properties
of simulated GCs depend on these parameters. 
Threshold gas densities for star formation ($\rho_{\rm th}$) can significantly
influence the final $M_{\rm sg}$ such that $M_{\rm sg}$ can be lower
for higher $\rho_{\rm th}$. If $\rho_{\rm th}$ is quite high
$\ge 10^5$, then SG formation is severely suppressed, which ends up
with SCs with small $M_{\rm sg}$ that can not be identified as genuine GCs.
This result implies that $\rho_{\rm th}$ could be similar between
FG and SG formation for self-enrichment scenarios to be viable. \\

(5) Formation of SG stars from AGB ejecta can last as long as
$\sim 10^8$ yr, because gaseous ejecta from AGB stars with different
masses thus different main-sequence lifetimes can be accreted onto the proto-GCs.
Accordingly, SNe from  SG stars formed earlier 
brow off the accumulated AGB ejecta so that star formation
can be severely suppressed.  This suppression of star formation
ends up with significantly smaller $M_{\rm sg}$ in GCs, which implies
that the mass budget problem is much more severe than ever thought
in the self-enrichment scenario of GC formation with multiple stellar
populations. Therefore
 the formation of GCs with $M_{\rm sg} \sim 10^5 {\rm M}_{\odot}$
requires a very small number fraction of high-mass stars
with $m_{\rm s} \ge 8 {\rm M}_{\odot}$ in SG star formation
(`top-light' IMFs).  Such suppression of massive star formation
in SG star formation was also pointed out by D10. \\

(6) The required top-light IMF in SG formation has some important
implications both  on the observed properties of GCs and young massive
SCs and on theoretical studies of star formation. 
First, even if secondary star formation is ongoing in
young massive SCs,  massive OB stars can not be
observed in the SCs owing to the lack of such massive stars.  
This may explain why recent observations of massive young SCs
did not detect signs of massive OB stars.
Second, the mass budget problem needs to be revisited, given that
the mass fraction of low-mass stars ($m_{\rm s} \le 0.9 M_{\odot}$, i.e., 
presently `alive' old stellar population) in SG subpopulation can be 
significantly larger for top-light IMF. 
Third, a mechanism for suppression of massive stars in dense stellar
systems  needs to be
theoretically understood. \\

(7) If top-light IMF is not possible in SG star formation, 
then $M_{\rm mc, th}$ can be quite large ($> 10^7 {\rm M}_{\odot}$)
in any self-enrichment scenario of GC formation owing to very low
$\epsilon_{\rm sf, sg}$ ($<0.1$).
Therefore the scenario
needs to explain how and why such  a large $M_{\rm mc, th}$ 
is possible in gas-rich dwarfs (or in other environments)
at high redshifts. If the scenario fails to
explain the physical origin of such high $M_{\rm mc, th}$,
then it would need to be discarded as a viable scenario for GC formation.
Thus, a possible IMF variation in star formation within dense stellar systems
will need to be investigated in theoretical studies of GC formation
based on self-enrichment scenarios.
\\

\section{Acknowledgment}
I (Kenji Bekki; KB) am   grateful to the referee  for  constructive and
useful comments that improved this paper.
Numerical simulations  reported here were carried out on
the two GPU clusters,  Pleiades and gSTAR kindly made available by International Center for radio astronomy research(ICRAR
) at  The University of Western Australia
and the Center for Astrophysics and Supercomputing
in the Swinburne University, respectively.
This research was supported by resources awarded under the Astronomy Australia Ltd'
s ASTAC scheme on Swinburne with support from the Australian government. gSTAR is funded b
y Swinburne and the Australian Government's
Education Investment Fund.
KB appreciate that Amanda Karakas provided stellar yields of AGB stars
for this research. 
KB acknowledges the financial support of the Australian Research Council
throughout the course of this work.

\appendix

\section{A way to set up initial fractal gaseous distribution of
MCs}

We generate an initial fractal distribution of gas in  a MC as follows.
First,  gas (SPH) particles with
the total particle number of $N_{\rm min}$ are distributed within
a sphere according to an adopted radial density profile
of the MC (i.e., $\rho_{\rm mc}(r) \propto r^{-1}$).
A random number generator is used in distributing these $N_{\rm min}$
particles.  This first
step is called Level 1 and the initial radius of the sphere
is denoted as $r_1$ for simplicity. Second, at $i$th particle's  position
($i=1,2, ... N_{\rm min}$),
new gas particles with the total number of
$N_{\rm min}$ are distributed within a sphere of $r_2$ using
the same radial profile
adopted in Level 1. 
The radius of the sphere in Level 2 is determined as follows:
\begin{equation}
r_2=\frac{ r_1 }{ f_{\rm div} },
\end{equation}
where $f_{\rm div}$ is a division factor, which is described as follows:
\begin{equation}
f_{\rm div}=N_{\rm min}^{1/D_3},
\end{equation}
where $D_3$ is the fractal dimension of the MC (as defined in the main
text). Accordingly, the large-scale 
particle distribution in Level 1 and the small-scale one  around 
$i$th gas particle are self-similar. This process is done for
each of  $N_{\rm min}$
particles  generated in the Level 1.
Thus, the total number of particles used in this  Level 2 is 
$N_{\rm min}^{ f_{\rm div} }$.

If the particle distribution around $i$th gas particle  in Level 2 is exactly
the same as the original particle
distribution in Level 1, then the final distribution of gas
becomes very artificial (`mathematical') one. In order to avoid
this, a random number generator is used each time when gas particle
distribution is generated for a given (adopted) radial distribution
of gas. By doing so, the final distribution of particles become more
natural in the present study. 
This process of generating a self-similar  particle distribution
is repeated in Level 3, 4, 5 etc
until the total number of particles becomes the adopted  number
of particles of a MC  in a simulation (i.e., $N_{\rm g} \sim 10^6$).
In the present study $N_{\rm min}$ is set to be 32, which ensures that
the initial distribution of gas particles in Level 1 can be a proper
representation of the adopted radial profile. If $N_{\rm min}$ is too small,
then the initial distribution is not so similar to the adopted profile.
On the other hand, $N_{\rm min}$ is large (e.g., 100), then the number 
of division becomes smaller.
We consider that the above number of 32 is appropriate for the present
investigation of GC formation within fractal MCs.

In order to give random motion of gas particles within a fractal MC,
we adopt the following model. We here consider that the total number of 
sub-groups of gas particle at the final Level of division (for the  fractal mass
distribution of a MC)
is $n_{\rm gr}$. These sub-groups have random motion within the MC characterized by
velocity dispersion $\sigma$.  Accordingly, $\sigma$ is determined as follows:
\begin{equation}
T_{\rm ran}=\frac{ 1 }{ 2 } \sum_{k=1}^{ n_{gr} } M_{\rm gr, \it k} \sigma^2,
\end{equation}
where $T_{\rm ran}$ is the total random (kinetic) energy of the MC
and $M_{\rm gr, \it k}$ is the mass of each sub-group.
Using a random number generator
and assuming an isotropic velociy dispersion, the 3D velocity of each sub-group
is given for the derived $\sigma$. 
In some models, a MC has initial rigid rotation with the amplitude of 
$\Omega$ (constant). 
Each sub-group's $\Omega$ is therefore determined as follows:
\begin{equation}
T_{\rm rot}=\frac{ 1}{ 2} \sum_{k=1}^{ n_{gr} } R_k^2 M_{\rm gr, \it k} \Omega,
\end{equation}
where $T_{\rm rot}$ is the total rotational  energy of the MC
and $R_k$ is the projected distance of $k$th sub-group from the MC's center.
Using the derived $\Omega$ and $R_k$,
rotational velocities of gas particles within each sub-group
are calculated.
Gas particles within a  sub-group
are assumed to 
have the same velocities in the present study.

\end{document}